\documentclass[aps, pre, superscriptaddress,notitlepage, amsmath]{revtex4-1}

\usepackage{graphicx}
\usepackage{hyperref}
\usepackage{amsmath,bm}
\usepackage[utf8]{inputenc}
\usepackage{color}
\usepackage[toc]{appendix}
\usepackage{ amssymb }
\usepackage{txfonts}
\usepackage[export]{adjustbox}
\newcommand{\avdata}[1]{ \left\langle{#1} \right\rangle_{\text{${data}$} } }
\newcommand{\avmodel}[1]{ \left\langle{#1} \right\rangle_{\text{$RBM$}} }

\newcommand{\av}[1]{ \left\langle{#1} \right\rangle}

\begin{document}
\title{`Place-cell' emergence and learning of invariant data with restricted Boltzmann machines: \\
breaking and dynamical restoration of continuous symmetries in the weight space}

\author{Moshir Harsh}
\affiliation{LPENS, Ecole Normale Sup\'erieure, CNRS UMR 8023 and PSL Research, 24 Rue Lhomond, 75231 Paris Cedex 05, France.}
\author{J\'er\^ome Tubiana}
\affiliation{Blavatnik School of Computer Science, Tel Aviv University, Israel}
\author{Simona Cocco}
\affiliation{LPENS, Ecole Normale Sup\'erieure, CNRS UMR 8023 and PSL Research, 24 Rue Lhomond, 75231 Paris Cedex 05, France.}
\author{R\'emi Monasson}
\affiliation{LPENS, Ecole Normale Sup\'erieure, CNRS UMR 8023 and PSL Research, 24 Rue Lhomond, 75231 Paris Cedex 05, France.}

\date{\today}

\begin{abstract}
Distributions of data or sensory stimuli often enjoy underlying invariances. How and to what extent those symmetries are captured by unsupervised learning methods is a relevant question in machine learning and in computational neuroscience. We study here, through a combination of numerical and analytical tools, the learning dynamics of Restricted Boltzmann Machines (RBM), a neural network paradigm for representation learning. As learning proceeds from a random configuration of the network weights, we show the existence of, and characterize a symmetry-breaking phenomenon, in which the latent variables acquire receptive fields focusing on limited parts of the invariant manifold supporting the data. The symmetry is restored at large learning times through the diffusion of the receptive field over the invariant manifold; hence, the RBM effectively spans a continuous attractor in the space of network weights. This symmetry-breaking phenomenon takes place only if the amount of data available for training exceeds some critical value, depending on the network size and the intensity of symmetry-induced correlations in the data; below this 'retarded-learning' threshold, the network weights are essentially noisy and overfit the data.
\end{abstract}

\maketitle

\section{Introduction}

Many high-dimensional inputs or data enjoy various kinds of low-dimensional invariances, which are at the basis of the so-called manifold hypothesis \cite{manifold}.  For instance, the pictures of somebody's face are related to each other through a set of continuous symmetries corresponding to the degrees of freedom characterizing the relative position of the camera (rotations, translations, changes of scales) as well as the internal deformations of the face (controlled by muscles). While well-understood symmetries can be explicitely taken care of through adequate procedures, e.g. convolutional networks, not all invariances may be known a priori. An interesting question is therefore if and how these residual symmetries affect the representations of the data achieved by learning models. 

This question does not arise solely in the context of machine learning, but is also of interest in computational neuroscience, where it is of crucial importance to understand how the statistical structure of input stimuli, be they visual, olfactive, auditory, tactile, ... shapes their encoding by sensory brain areas and their processing by higher cortical regions. Information theory provides a mathematical framework to answer this question \cite{Laughlin}, and was  applied, in the case of linear models of neurons, to a variety of situations, including the prediction of the receptive fields of retinal ganglion cells \cite{atick1992could}, the determination of cone fractions in the human retina \cite{Vijay} or the efficient representation of odor-variable environments \cite{Tibi}. In the case of natural images, which enjoy approximate translational and rotational invariances, non-linear learning rules resulting from adequate modification of Oja's dynamics \cite{gerstner} or sparse-representation learning procedures \cite{sparse} produce local edge detectors, such as do independent component analysis  \cite{hyvarinen2000independent}. These detectors bear strong similarities with the neural receptive fields measured in the visual cortex (V1 area) in mammals.

It is therefore natural to wonder whether the existence of localized receptive fields is a general feature to be expected from representations of invariant distributions of inputs. Gardner's theory of optimal learning for single-layer neural network (perceptron) predicts that spatially correlated patterns, {\em e.g.} drawn from a translationally-invariant distribution, lead to a localized profile of weights \cite{monasson93}. Further supporting evidence was recently brought by several works, focusing on the production of such receptive fields in the context of unsupervised learning. Learning of symmetric data with similarity-preserving representations \cite{sengupta} or with auto-encoders \cite{benna} both led to localized receptive fields tiling the underlying manifold, in striking analogy with place cells and spatial maps in the hippocampus. In turn, such high-dimensional place-cell-like representations have putative functional advantages: they can be efficiently and accurately learned by recurrent neural networks, and thus allow for the storage and retrieval of multiple cognitive low-dimensional maps \cite{battista19}.

The present work is an additional effort to investigate this issue in a highly simplified and idealized framework of unsupervised learning, where both the data distribution and the machine are under full control. Similarly to previous studies \cite{mehtaschwab,ringel}, we consider synthetic data with controlled invariances generated by standard models instatistical physics, such as the Ising and XY models. These data are then used to train Restricted Boltzmann Machines (RBM), a simple albeit powerful framework for representation learning, where a layer of hidden (latent) units account for the correlation structure in the data configurations. We show how the receptive fields of the hidden units undergo a symmetry-breaking transition in the space of couplings: units individually cover localized regions of the input space, but concur to tile the space as best as possible, in much the same way as hippocampal place cells do. This symmetry breaking is dynamically restored if we let the training algorithm run for very long times (well beyond the training time needed to saturate the log-likelihood of the test set): while keeping their localized shape, the center of the receptive/place fields diffuses along the input space, effectively ensuring the invariance of the learned distribution. We also show that this symmetry-breaking phenomenon requires a minimum number of data, an illustration of the general phenomenon of retarded learning \cite{watkin}, also encountered in random matrix theory in the context of the so-called spiked covariance model \cite{reimann,bbp}. 

Our paper is organized as follows. RBM and their learning algorithms are introduced in Section~II. We consider the case of a data distrbution with a single invariance in Section~III, and with two symmetries in Section~IV. A detailed theoretical analysis of the learning dynamics and of the receptive field emerging through the symmetry-breaking transition can be found in Section~V. Conclusions and speculative connections with experiments in neuroscience are proposed in Section~VI.

\section{Restricted Boltzmann Machines}

\subsection{Definition and log-likelihood}

\begin{figure}[h]
\begin{center}
\includegraphics[scale=0.25]{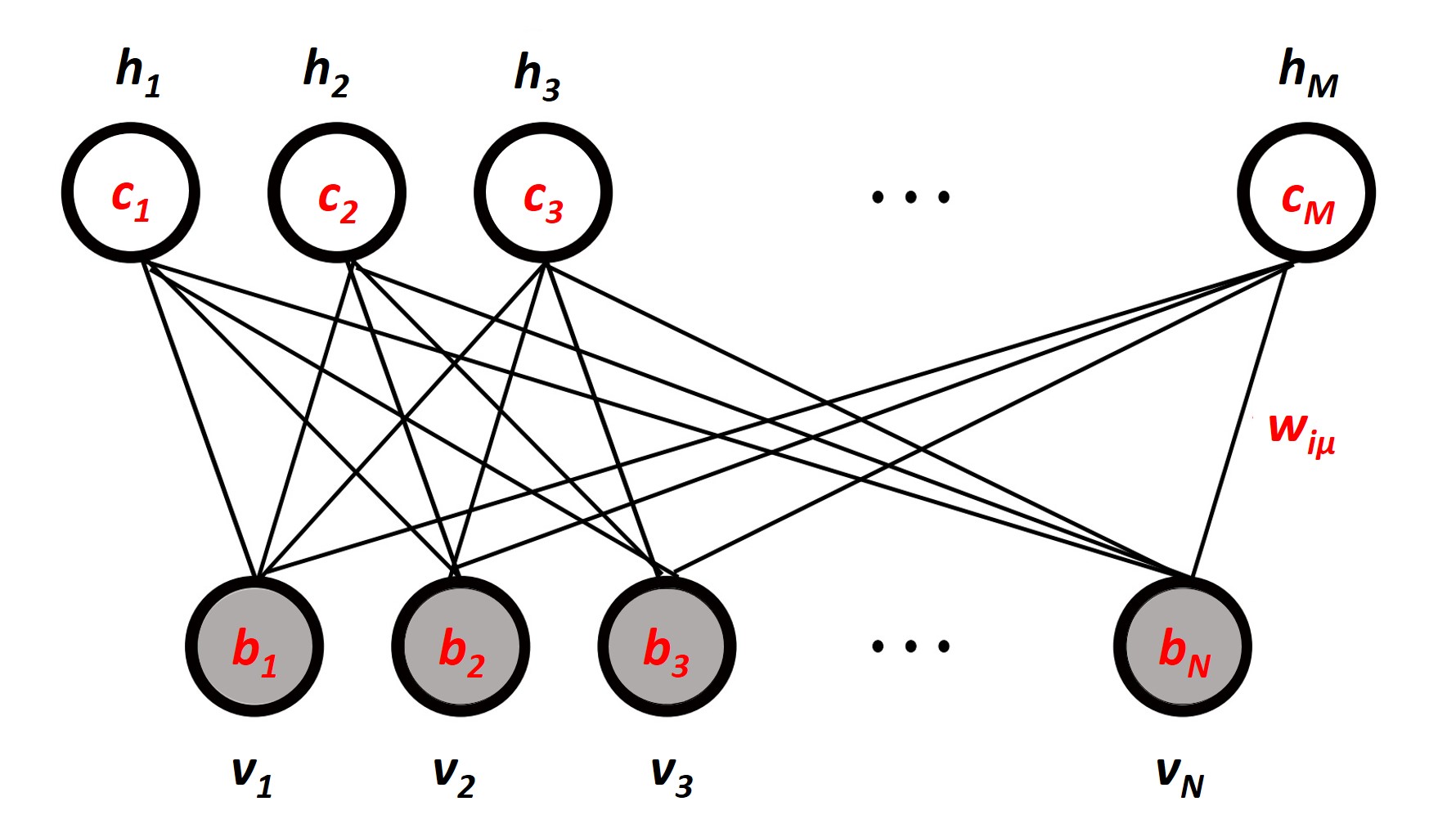}
\caption{The two-layer structure of RBM, with weigts $w_{i\mu}$ connecting $N$ visible units $v_i$ to $M$ hidden units $h_\mu$. These binary-valued units are subjected to local fields, called $b_i$ and $c_\mu$ for, respectively, the visible and hidden layers.}
\label{fig_rbm}
\end{center}
\end{figure}

A Restricted Boltzmann Machine (RBM) is a bipartite, undirected stochastic neural network with two layers, see Fig.~ \ref{fig_rbm}:

\begin{itemize}
\item the {\em visible} layer includes $N$ units $v_i$, $i=1,..., N$, which carry the configurations of data. For simplicity, we assume here that visible units take binary values, $v_i=\pm 1$.
\item the {\em hidden} layer includes $M$ units $h_\mu$, $\mu=1,..., M$, on which are expressed the representations of the data configurations. Hidden, or latent variables $h_\mu$ can take real or binary values.
\end{itemize}

The model is formally defined by a Gibbs probability distribution over the sets of visible (${\bf v}$) and hidden (${\bf h}$) variable configuration:

\begin{equation} 
p(\textbf{v},\textbf{h}) = \frac{1}{\cal Z}e^{-E(\textbf{v},\textbf{h})}\ , \quad \text{where} \quad {\cal Z} = \sum_{ {\bf v}} \int d{\bf h} \, e^{-E({\bf v},{\bf h})}
\end{equation}
is the partition function, such that $p$ is normalized to unity, and the energy function $E(\textbf{v,h})$ is given by
\begin{equation}\label{defE}
	E(\textbf{v},\textbf{h}) = -\sum_{i=1}^{N}\sum_{\mu=1}^{M} w_{i\mu}\, v_i\, h_\mu -\sum_{i=1}^{N} b_i \, v_i+ \sum_{\mu=1}^{M} \mathcal{U}_\mu( \, h_\mu) \ .
\end{equation}
In the formula above, $w_{i\mu}$ is the real-valued weight (coupling) connecting the hidden unit $h_{\mu}$ and the visible unit $v_{i}$, $b_i$ are real-valued bias terms, also called fields and $\mathcal{U}_\mu$ are the hidden unit potentials. We consider two possible choices for $\mathcal{U}_\mu$:
\begin{itemize}
\item For binary ($\pm 1$) valued hidden units, a regular field term $\mathcal{U}_\mu (h_\mu) = - c_\mu h_\mu $ similar to the visible units. In that case, Eqn.~\ref{defE} is a special case of Ising distribution, with only couplings between units belonging to different layers.

\item For real valued hidden units, the symmetric double well potential $\mathcal{U}_\mu(h_\mu) =  \frac{1}{2} h_\mu^2 + \theta_\mu\, | h_\mu |$. For $\theta_\mu=0$, the potential is quadratic and the corresponding variable is Gaussian and for $\theta_\mu<0$ the potential has two minimas at $\pm \theta_\mu$; this choice of potential effectively interpolates between Gaussian ($\theta_\mu=0$) and binary ($\theta_\mu \rightarrow -\infty$) hidden units \cite{barra2017phase}.
\end{itemize}

Due to the absence of connections between the units within a layer, the conditional probability of hidden units given the visible units factorizes as follows:
\begin{equation} \label{conditional_sampling}
p( {\bf h}|{\bf v}) = \prod_{\mu=1}^{M}p \left( h_\mu| I_\mu ({\bf v}) \right)\ ,
\end{equation}
where $I_\mu({\bf v}) = \sum_i w_{i\mu} v_i$ is the total input received from the visible layer by hidden unit $\mu$ in the absence of fields on visible units, and $p(h_\mu | I) \propto e^{\mathcal{U}_\mu(h_\mu) + h_\mu\, I}$. Therefore, sampling from the conditional distribution is simply done by first computing the hidden layer inputs $I_\mu$, then sampling independently each hidden unit given its input according to its hidden unit potential. Similarly, the average activity of a hidden unit given the visible units, $\av{ h_\mu | {\bf v} }$, is a non-linear function of the input $I_\mu({\bf v})$ ; for binary hidden units, we have $\av{h_\mu | {\bf v} } = \tanh (\sum_i w_{i\mu} v_i + c_\mu)$. Therefore, RBM can be viewed as linear-nonlinear model similar to other feature extraction methods such as Independent Component Analysis. Symmetric formulas can be written for the conditional probability of visible units given the hidden units.

In addition, the marginal distribution over the visible units $p({\bf v})$ can be written in closed form:
\begin{equation}\label{marginal}
p(\textbf{v}) =  \int d{\bf h} \, p(\textbf{v},\textbf{h}) = \frac{1}{\cal Z}\, e^{\sum_{i=1}^N b_i v_i} \prod_{\mu=1}^N \int  dh_\mu \, e^{-\mathcal{U_\mu}(h_\mu) + h_\mu \, I_\mu({\bf v}) } = \frac{1}{\cal Z} \exp \bigg( \underbrace{\sum_{i=1}^N b_i v_i + \sum_{\mu=1}^M \Gamma_\mu\big(I_\mu ({\bf v})\big)}_{-E_\text{eff}({\bf v})} \bigg) \nonumber \ ,
\end{equation}
where $\Gamma_\mu(I) = \log \int dh \, e^{-{\cal U}_\mu(h) + h \,I} $ is the cumulant generative function, or log Laplace transform, associated to the potential ${\cal U}_\mu$; for binary hidden units, $\Gamma_\mu(I) = \log 2 \cosh (I+c_\mu)$. Note that by construction, $\Gamma_\mu'(I_\mu)$ is the average value of the hidden unit given its input $I_\mu$; therefore the hidden unit potential determines the transfer function of the hidden unit. Importantly, although the joint distribution is pairwise, the marginal distribution is not in general as $\Gamma_\mu$ functions are not quadratic. Therefore, RBM generate effective high-order interactions between the units $v_i$, and are capable of expressing complex measures over the visible configurations \cite{le2008representational,tubiana2017emergence}.

\subsection{Training algorithm}

Training the RBM is the process of fitting the parameters $\boldsymbol\Theta= \{ w_{i\mu},b_i,c_\mu/\theta_\mu \}$ to maximize the average log-likelihood of the $S$ data items $\textbf{v}^{data}$ assumed to be independently drawn from $p(\textbf{v})$. While this may be done with the gradient ascent method, calculating the likelihood is computationally intensive as it requires evaluating the partition function, and sampling methods like Markov Chain Monte Carlo (MCMC) in the form of Gibbs sampling are used.
	
\subsubsection{Gradient of log-likelihood}
	
For the model with parameters $\boldsymbol\Theta$, the log-likelihood of a single training example $\textbf{v}^{data}$ is
\begin{equation}
	\log \mathcal{L}(\textbf{v}^{data} | \boldsymbol{\Theta} ) = \log p\left( {\bf v}^{data}\right) = - E_{\text{eff}}({\bf v}^{data}) - \log {\cal Z} = - E_{\text{eff}}({\bf v}^{data}) - \log \left[ \sum_{\bf v} e^{- E_{\text{eff}}({\bf v}) } \right]\ .
\end{equation}
Taking the partial derivative with respect to $\Theta$ gives
\begin{equation}
\frac{\partial \log \mathcal{L} (\textbf{v}^{data} | \boldsymbol{\Theta} )}{\partial \boldsymbol{\Theta} } = - \frac{\partial E_{\text{eff}}({\bf v}^{data} )}{\partial \boldsymbol{\Theta} } + \frac{1}{\cal Z} \frac{\partial \cal Z}{\partial \boldsymbol{\Theta} } = - \frac{\partial E_{\text{eff}}({\bf v}^{data} )}{\partial \boldsymbol{\Theta}} + \avmodel{ \frac{\partial E_{\text{eff}}({\bf v} )}{\partial \boldsymbol{\Theta}} } \ ,
\end{equation}
where $\avmodel{(.)} = \frac{1}{\cal Z} \sum_{\bf v} e^{-E_{\text{eff}}({\bf v})} (.) $ denotes the average according to the marginal distribution over the visible units with parameter values $\boldsymbol{\Theta}$. 

In particular, for the weights $w_{i\mu}$, we have according to (\ref{marginal}), $\frac{\partial E_{\text eff}(\textbf{v})}{\partial w_{i\mu}}=-v_i\ \Gamma_\mu'\left(I_\mu({\bf v}) \right) \equiv -v_i \av{h_\mu | {\bf v}}$. The gradient of the total log-likelihood is then

\begin{equation}\label{gradL1}
\frac{\partial \avdata{ \log \mathcal{L} (\textbf{v}^{data} | \boldsymbol{\Theta} ) } }{\partial w_{i\mu}} = \avdata{ v^{data}_i \av{ h_\mu | {\bf v}^{data} } } - \avmodel{ v_i \av{ h_\mu | {\bf v} } } \ . 
\end{equation}
Equation (\ref{gradL1}) is an example of moment-matching condition, as it imposes that the correlation between the variables $v_i$ and $h_\mu$ computed from the data coincides with its counterpart defined by the RBM model distribution $p(\mathbf{v},\textbf{h})$. The gradients of ${\cal L}$ over $b_i$ and $c_\mu$ lead to similar moment-matching conditions for, respectively, the average values of $v_i$ and of $h_\mu$.
	
	\subsubsection{Approximating the log-likelihood gradient}
	
In the gradient of the log-likelihood of Eqn.~(\ref{gradL1}), the model-distribution moment is not computationally tractable, as it requires to sum over all values of the visible and the hidden variables. In practice, an approximate value for this term is obtained by Markov Chain Monte Carlo (MCMC) methods. The Markov Chain is defined by repeated iterations of Gibbs sampling, which consists in sampling $\textbf{h}$ from $\textbf{v}$ and $\textbf{v}$ from $\textbf{h}$ using Eqn. \ref{conditional_sampling}. 
In principle, one should run a full MCMC simulation at each gradient step, but this is computationally prohibitive. For our RBM training we use the \textit{Persistent Contrastive Divergence (PCD)} algorithm \cite{tieleman2008training}: Markov Chains are initialized at the beginning of the training and updated with only a few Gibbs Monte Carlo steps between each evaluation of the gradient, see \cite{fischer2015} for a more detailed review. This approximation works very well for the data distribution studied here because they are in a paramagnetic phase (= monomodal), hence the Markov Chains mix very rapidly.

\subsubsection{Stochastic Optimization}

The RBM is trained using Stochastic Gradient Ascent (SGA), the golden standard for neural network optimization. SGA is a variant of ordinary gradient ascent where at each step, only a small subset of the data set (the minibatch), of size $B \sim 10-100$ examples, is used to evaluate the average log-likelihood, see Eqn~\ref{equpdate} where $\nu$ is the learning rate which dictates how much to change the parameter in the direction of the steepest gradient. The dataset is divided into $S/B$ mini-batches $Batch(t)$, and for each epoch of training $t$, we perform one SGA update for each mini-batch. An epoch consists of using all the subsets for the update such that each data sample is used once. After every epoch the subsets are again drawn randomly. Several dozens of epochs are usually required to reach convergence.

	\begin{equation}\label{equpdate}
	\boldsymbol{\Theta}^{t+1} = \boldsymbol{\Theta}^{t} + \nu \left[ \frac{1}{B} \sum_{b\  \in\ \text{Batch}(t)} \nabla_{\boldsymbol{\Theta}} \log \mathcal{L} (\textbf{v}^{data,b} | \boldsymbol{\Theta} ) \right]
	\end{equation}

Compared to ordinary gradient ascent, SGA serves several purposes. First and foremost, its computational cost is significantly reduced as only a small batch is used per update; yet the update is usually reliable thanks to data redundancy. Second, the stochastic evaluation of the gradient introduces noise in the learning process. This prevents the dynamic from getting trapped in local maxima, which is crucial for non-convex optimization landscapes, and it also directs the dynamics toward minima with wider basins of attraction \cite{jastrzkebski2017three}. It has been argued that the later effect contributes in improving generalization performance \cite{hochreiter1997flat,keskar2016large,chaudhari2016entropy}. Though the convergence rate of SGA has a slower asymptotic rate than ordinary gradient descent, it often does not matter in practice for finite data sets, as the performance on the test set usually does not improve anymore once the asymptotic regime is reached \cite{bottou2008tradeoffs}.

The noise level of the SGD is directly related to the batch size and learning rates parameters, see for instance \cite{smith2017bayesian}. Briefly speaking, assuming i.i.d. and infinite number of samples, the SGA parameter increment has mean value $ \nu \nabla_{\boldsymbol{\Theta}} \avdata{ \log \mathcal{L} (\textbf{v}^{data} | \boldsymbol{\Theta} ) }$, and variance proportional to $\nu^2/B$; in the large $B$ limit it is also Gaussian distributed according to the central limit theorem. In comparison, the increments of a continuous time Langevin equation with energy landscape $E=-\avdata{ \log \mathcal{L} (\textbf{v}^{data} | \boldsymbol{\Theta} ) }$ and noise covariance matrix $\propto \sigma_{SGA}^2$, integrated over a time step $\nu$ has the same mean value and a covariance proportional to $\sigma_{SGA}^2 \nu$. Identifying both noises gives the following scaling law for the SGA noise, $\sigma_{SGA} \propto \sqrt{\frac{\nu}{B}}$. Reducing the learning rate and increasing the batch size therefore decrease the noise level, and vice-versa. In all our experiments, both learning rates and batch sizes are kept fixed throughout a training session.

\section{Learning Data with a Single Invariance}

\subsection{Data distribution: Ising model}

Our first toy distribution for data $\textbf{v}^{data}$ is the celebrated one-dimensional ising model from statistical physics. Here each $v_i$ is a spin which can either be up or down, that is can take only $\pm 1$ binary values. The corresponding joint probability distribution of the visible units reads
	\begin{equation}{\label{p_ising1D}}
	p_{data}(v_1,v_2,...,v_N) = \frac{1}{Z_{ising}} e^{\, \beta\, \sum_{i=1}^{N }v_i \, v_{i+1}}
	\end{equation}
where the partition function normalizes this probability over the $2^N$ visible configurations, and $\beta >0$ is referred to as the inverse temperature. We enforce periodic boundary conditions through $v_{N+1} \equiv v_1$.

As is well known, under distribution (\ref{p_ising1D}), all visible units $v_i$ have average values equal to zero, and the correlation function decays exponentially with the distance separating the corresponding units on the ring,
\begin{equation}\label{isinghj}
\left<v_i v_j\right> = \sum _{\textbf{v}} p_{data}(v_1,v_2,...,v_N) \, v_i \, v_j = e^{-|i-j|/\xi} \ , \quad \text{where} \quad  \xi = \frac 1{\ln \coth \beta} 
\end{equation}
is the correlation length. The above expression for the correlation holds when $N \gg \xi$.

\begin{figure}[h]
\hskip 1cm
  \begin{minipage}[c]{0.3\textwidth}
    \includegraphics[width=\textwidth]{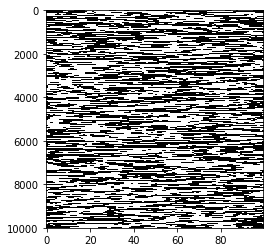}
  \end{minipage}\hfill
  \begin{minipage}[c]{0.55\textwidth}
    \caption{10,000 one-dimensional Ising model configurations with 100 spins each, sampled from distribution (\ref{p_ising1D}) at inverse temperature $\beta=1$ and with periodic boundary conditions. Black and white dots represent units equal to, respectively, $+1$ and $-1$. The correlation length $\xi$ may be though of as the typical length of black or white contiguous regions along the horizontal direction. Here, $\beta=1$, which corresponds to $\xi\simeq 3.7$.  } 
    \label{ising1D_config}
  \end{minipage}
\end{figure}

Formula (\ref{p_ising1D}) defines a simple example of invariant distribution under the set of translations (or, better, rotations) on the $N$-site ring. More precisely, for any integer $k$, we have
\begin{equation}\label{isingsym}
p_{data}(v_1,v_2,...,v_N) = p_{data}(v_{k+1},v_{k+2},...,v_{k+N}) \ ,
\end{equation} 
where $i+k$ is to be intended modulo $N$.
Figure~\ref{ising1D_config} shows a number of configurations, drawn independently and at random from this probability distribution using the Gibbs sampling algorithm. 

\subsection{Initial learning and emergence of place cells}

\subsubsection {Case of a single hidden unit }

\begin{figure}[b]
		\centering
		\textbf{(a)}
		\includegraphics[width = 0.45\linewidth,valign=t]{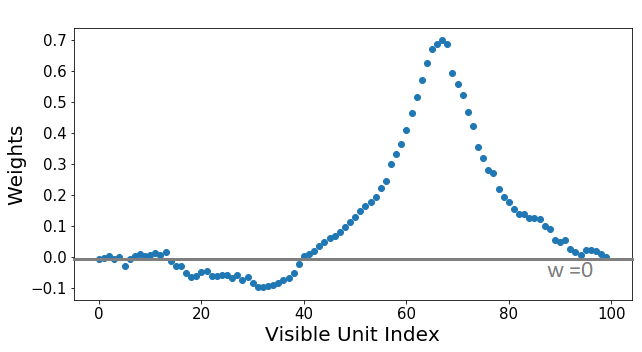}
		\hskip .5cm 
 \textbf{(b)}   \includegraphics[width=.45\textwidth,valign=t]{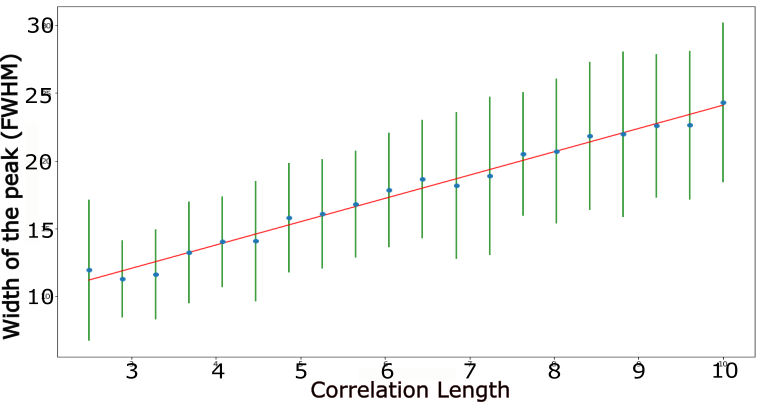}
   \caption{ {\bf (a)} Structure of weights learned by a RBM trained weights after 100 epochs from 10,000 data configurations  of the one-dimensional Ising model of size $N=100$. Training parameters: rate $\nu=0.001$, batch size $S_{batch}=10$, PCD-20 learning. One observes the emergence of a peaked structure in the weights, centered around site $i\simeq 68$. Note the small fluctuations in the tails (small $w_{i1}$), due to the finite (but large) number of data. {\bf (b)} Width of the peak in the weight space as a function of the correlation length of the Ising model, $\xi$. We trained our RBM with one hidden unit 25 times on data generated at different temperatures, $\beta$, and then calculated the average peak width and the standard deviation (error bars) over the different samples. The width was calculated by fitting a cubic spline with one knot to the profile: $y(i)=\left\lbrace w_i-\max(w_i)/2\right\rbrace$, where $i$ is the site index; The roots of this spline were then determined numerically, and the width was defined as the modulus of the difference between the roots. This procedure reliably finds the Full Width at Half Maximum (FWHM). A linear fit (red line) of the form $y=a\, x+b$ shows that the width of the place (receptive) field of the only hidden unit is proportional to the relevant characteristic length in the data. Notice that the intercept ($b$) is non zero, in agreement with the theoretical findings of Section \ref{betazero} in the  $\beta\rightarrow 0$ limit. } 
      \label{rbm1Dising}
\end{figure}

	First we train the RBM with only $M=1$ hidden unit, and $N=100$ visible units. Such a limited machine is, of course, not expected to reproduce accurately the Ising model distribution underlying the data. However, this is an interesting limit case to study how the RBM can make the most of its single set of weight attached to the unit. We use a large number of data configurations for training, which makes our distribution approximately invariant under rotations on the ring.
	
		 We initialize the weights $w_{i1}$ with small amplitude Gaussian random values; since the data are symmetric, we further impose $b_i=c_\mu =0\ \forall i,\mu$. The results of the training phase after 100 epochs, {\em i.e.} the weights $w_{i1}$ are shown in Fig.~\ref{rbm1Dising}(a). We observe that the weights are not uniform as could have been naively expected from rotational invariance, but focus on a limited portion (place) of the $N$-site ring. The position of the peak depends on the initial conditions for the weights;  it may also be influenced by the small irregularities in the data set coming from the finite number of training configurations.
		 
		 To understand what determines the width of the weight peak, we train different RBMs with data at different inverse temperatures $\beta$, and calculate their average peak widths over multiple runs. We plot the peak width as a function of the correlation length $\xi$ in Fig.~\ref{rbm1Dising}(b). We observe that the peak width scales proportionally to  $\xi$. Interestingly, despite its very limited expression power, our  single-unit RBM has correctly learned to coarse grain the visible unit configurations on the relevant scale length in the data, $\xi$. Having wider receptive, or place fields would not be as much as informative. For instance, with a set of uniform weights $w_{i1}=w$, the hidden unit would simply estimate the average magnetization of (mean value of all visible units in) the data configurations, which are all equal to zero up to fluctuations of the order of $\pm N^{-1/2}$, and would completely miss the correlated structure of the data. Conversely, more narrow place fields would have lower signal-to-noise ratios: the strong correlations of visible units over the length $\xi$ allows one to reliably estimate the local magnetization and the correlation structure on this scale. 		 


\begin{figure}[b]
		\centering
		\textbf{(a)}
		\includegraphics[width = 0.45\linewidth, valign=t]{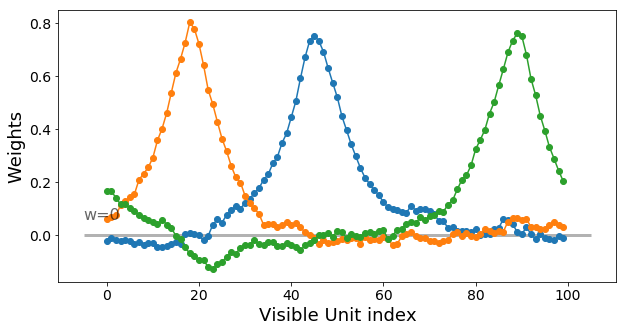}
		\hskip .3cm 
		\textbf{(c)}
		\includegraphics[width = 0.45\linewidth, valign=t]{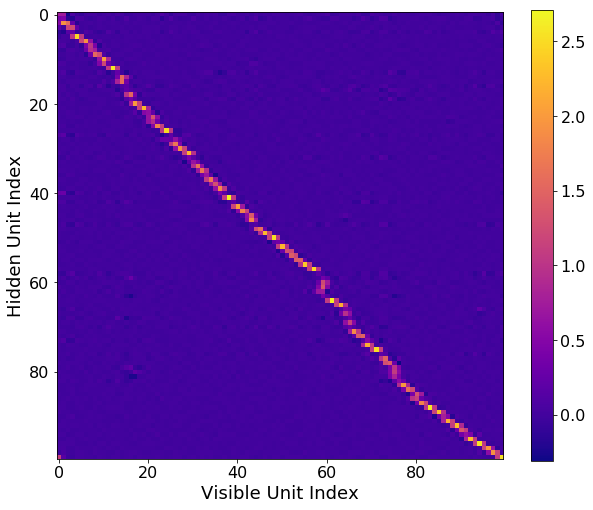}
\vskip -2.5cm \hskip -8.5cm
		\textbf{(c)}
		\includegraphics[width = 0.45\linewidth, valign=t]{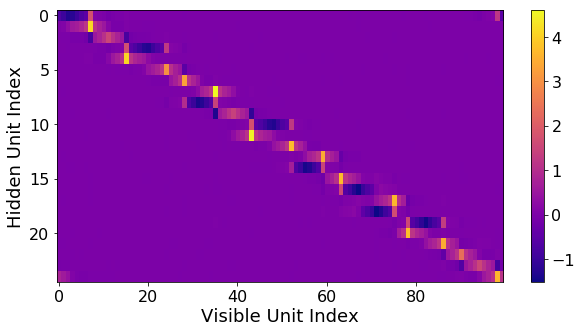}
		\caption{ {\bf (a)}
		Same as Fig.~\ref{rbm1Dising}(a), but with a RBM having $M=3$ hidden units.The weights attached to the same hidden units are shown with the same color. The receptive fields (peaks) for the three hidden units are roughly equally separated from each other. Note that due to the invariance of the probability distributions of the RBM under $h_\mu\to -h_\mu, w_{i\mu}\to -w_{i\mu}$ (when $c_\mu=0$), the overall sign of the weights attached to the same hidden unit does not matter.
		{\bf (b) \& (c)} Same as Fig.~\ref{rbm1Dising}(a), but with a RBM having, respectively $M=25$ (\textbf{a}) and $M=100$ (\textbf{c}) hidden units. The color codes show the intensity of the weights $w_{i\mu}$ as a function of the hidden ($\mu$, $y$-axis) and visible ($i$, $x$-axis) unit labels. The hidden units have been arranged according to the centre of their respective receptive field.}
		\label{rbm1Dising3units}
\end{figure}

\subsubsection {Case of multiple hidden units}\label{sec_M}
	
We next show results obtained when training RBM with $M=3$ hidden units on the same data. Figure~\ref{rbm1Dising3units} shows that each one of the three sets of weights have roughly the same peaked structure (same width) as in the $M=1$ case, but the peaks are centered at different places along the ring. The roughly equal distance between successive peaks shows the existence of an effective repulsion between the weights of any two hidden units. This phenomenon is easy to understand on intuitive grounds: having very overlapping place fields produces highly redundant hidden units,and would not help capturing the spatial correlation in the data spreading over the entire ring.

Training of RBMs with a large number of hidden units shows the same pattern of production of place fields attached to different hidden units, covering in a approximately uniform way the visible space (ring), see Fig.~\ref{rbm1Dising3units}(c) in the case of $M=100$ hidden units. The only notable difference is that the width of the place fields shrinks as $M$ gets very large. This happens when $M\xi \gg N$, {\em i.e.} when the single-hidden-unit peaks would start to largely overlap.

\subsection{Long-time learning and restoration of invariance through place-field diffusion}

\begin{figure}[h]
		\centering
		\textbf{(a)}
		\includegraphics[width = 0.9\linewidth, valign=t]{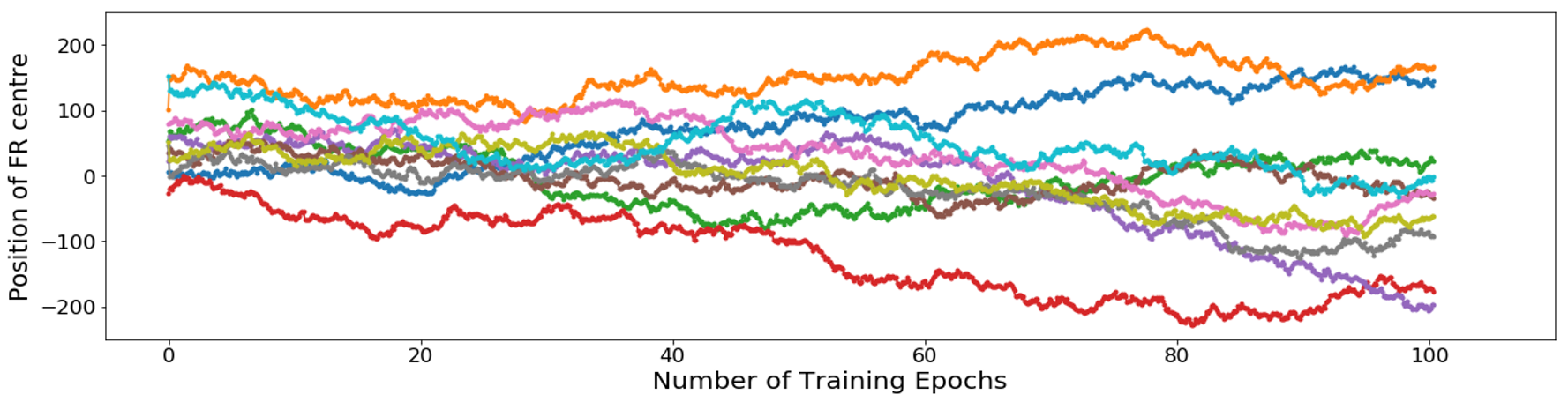}
		\newline
		 \begin{minipage}[c]{0.45\textwidth}
    \includegraphics[width=\textwidth]{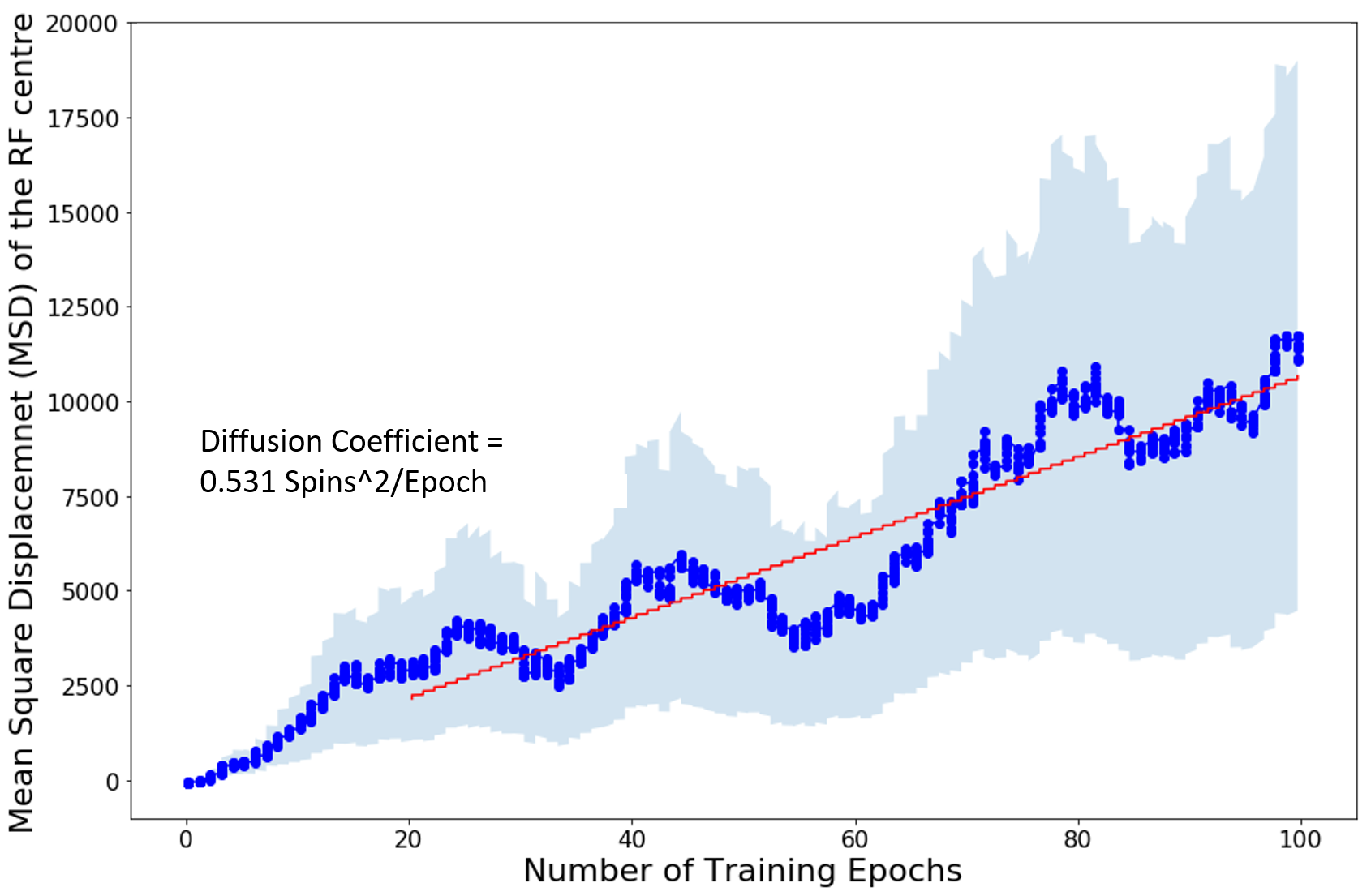}
  \end{minipage}\hfill
  \begin{minipage}[c]{0.5\textwidth}
    \caption{ {\bf (a)} Trajectories of the peak of the receptive field for a RBM with one hidden unit vs. number of epochs of training. Each trajectory corresponds  to a run of the training procedure starting from a random initial condition for the weights.  Training parameters: rate $\nu=0.1$, PCD-20 training and batch size $S_{batch}=100$, 1 million configurations of the one-dimensional Ising  model at  $\beta=1$. Trajectories are corrected for periodic boundary conditions - when the peak crosses the boundary, we add or subtract the size $N$ of the system. 
			{\bf (b)} Mean Square Displacement (MSD) of the peak of the receptive field vs. number $t$ of epochs of training. Time $t=0$ corresponds to the beginning of training. The red line is a linear fit $= 2\,D\,t$ obtained after removing the transitory period of the first 20 epochs of training. Results obtained from the trajectories shown in (a).
} 
\label{smooth1D1}
  \end{minipage}
	\end{figure}

We now let the training dynamics evolve for a much larger number of epochs. In the case of a RBM with one hidden unit only, the weight vector shows the overall peak structure of Fig.~\ref{rbm1Dising}(a) at all times (after a short initial transient during which the localized peak emerges). However, the location of the peak may change on very long training time scales. Figure~\ref{smooth1D1}(a) shows ten trajectories of the center of the peak corresponding to ten random initialization of the weights (equal to small values drawn independently and randomly). We observe that the centers of the peaks undergo a seemingly random motion. When the number of data items used for training is very large (to erase any tiny non-homogeneity in the empirical distribution), this random motion looks like pure diffusion.

	\begin{figure}[h]
		\centering
		\includegraphics[width = 0.8\linewidth]{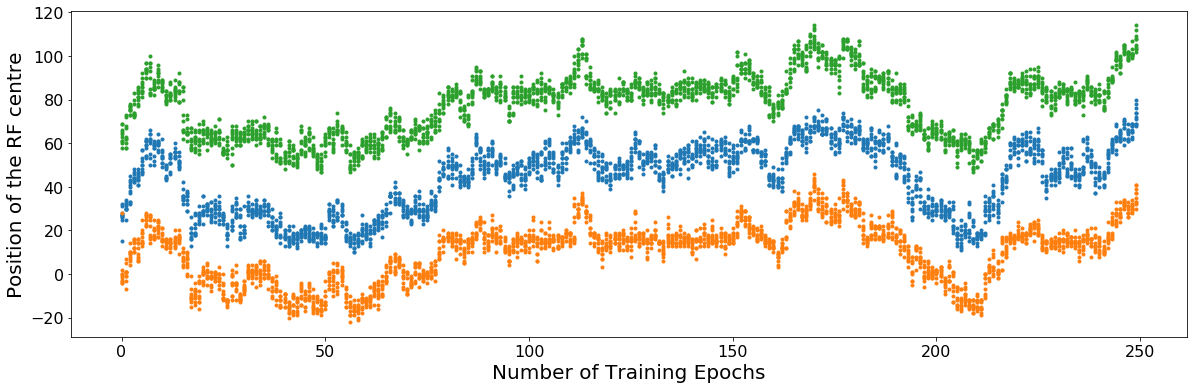}
		\caption{Locations of the peaks of the receptive fields for a RBM with $M=3$ hidden units as functions of the number of epochs of training.  Same training parameters as in Fig.~\ref{smooth1D1}. The trajectories are corrected for periodic boundary conditions - when the peak crosses the boundary, we add or subtract the size $N$ of the system. Multiple trajectories are repeated runs of the training procedure, starting from different initial random conditions for the weights. The  correlated motion of the place fields is a clear signature of the presence of repulsive interactions between the corresponding weight vectors. }
		\label{smooth1D_3u}
	\end{figure}

Figure~\ref{smooth1D1}(b) shows that the Mean Square Displacement (MSD) of the peak center grows roughly linearly with the training time (number of epochs), which defines the effective diffusion coefficient of the weight peak. For intermediate numbers of data items, diffusion is activated: due to the inhomogeneities in the empirical data distribution, some places along the ring are preferred, and have a tendency to trap the weight peak for some time.

Repeating the same analysis for a RBM with $M=3$ hidden units allows us to observe the  diffusion of the three peak centers, see Fig.~\ref{smooth1D_3u}. We see that the motions of these centers are coupled to maintain a constant distance between each other. This is a clear signature of the effective repulsion between the hidden-unit weight vectors already discussed in Section \ref{sec_M}.

\subsection{Case of few data: Retarded learning transition}

The emergence of a pronounced peak in the weight vector attached to a hidden unit reported above takes place only if the number of data items are sufficiently large. For very few data, the RBM weights do not show any clear spatial structure and seem to overfit the data. Similarly, for a fixed number of data samples, a transition is observed between the overfitting and spatially-structured regimes as the correlation length $\xi$ (or the inverse temperature $\beta$), that is, the spatial signal in the data is increased. To distinguish these two regimes, we introduce the empirical order parameter
\begin{equation} \label{op1}
W=\left| \sum _{i=1}^N w_{i1} \right| \ ,
\end{equation}
which is expected to be large when place fields emerge and the weights are spatially structured, and much smaller (and vanishingly small in the large--$N$ limit) in the overfitting regime. 

\begin{figure}[h]
		\centering
		\textbf{(a)}
		\includegraphics[width=0.45\linewidth,valign=t]{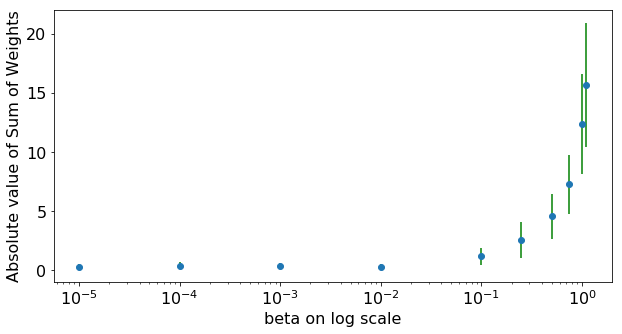} 
		\hskip .5cm
		\textbf{(b)} 	
		\includegraphics[width=0.45\linewidth,valign=t]{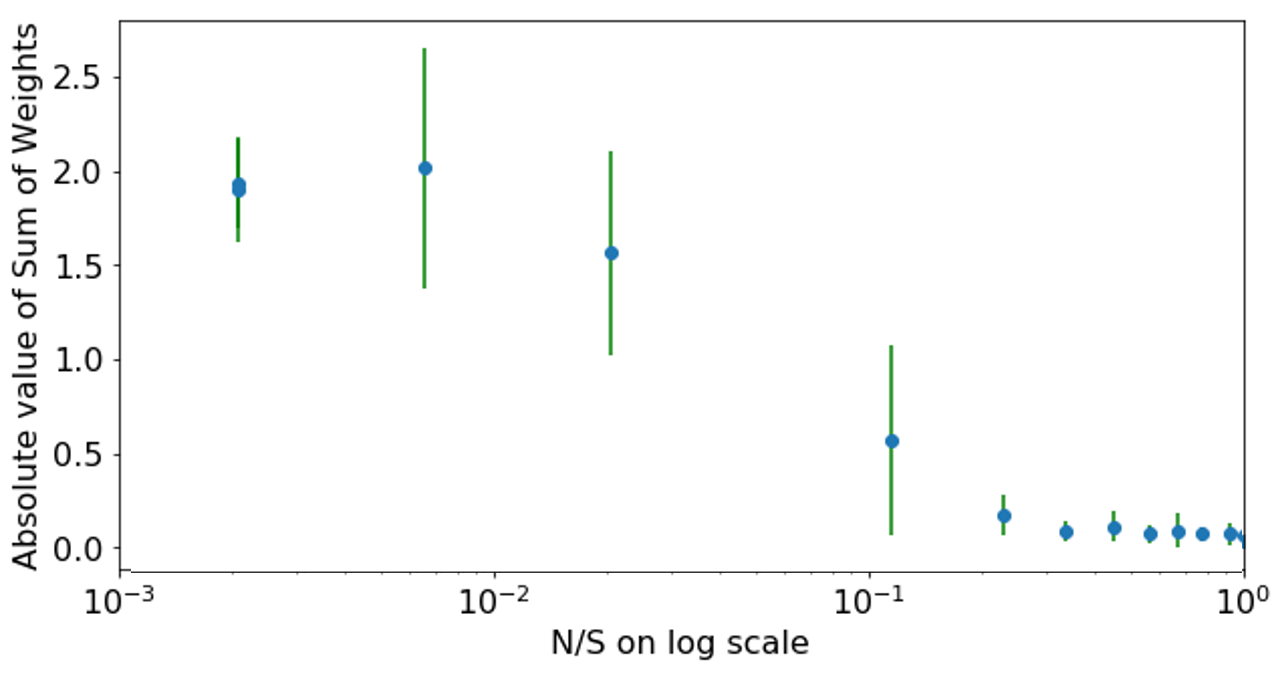} 		
		\caption{
		{\bf (a)}  Sum of weights $W$ after training a one-hidden-unit RBM on 10,000 configurations of the one-dimensional Ising model at different $\beta$. Here, there are $N=100$ visible units, {\em i.e.} noise ratio $r=N/S = 0.01$, other RBM training parameters are: $\nu=0.001$, $S_{batch}=10$, trained for 250 epochs with PCD-20. The error bars are calculated over 10 runs of the same training. 
		{\bf (b)} Sum of weights $W$ after training a one-hidden-unit RBM on configurations of the one-dimensional Ising model at $\beta=0.5$ as a function of the noise ratio  $r=N/S$. Here, there are $N=20$ visible units. All other parameters of training are same as (a) . The error bars are calculated over 10 runs of the same training. }
		\label{ising1D_RBMtransition}
	\end{figure}

Figure~\ref{ising1D_RBMtransition}(a) shows the value of the order parameter $W$ as a function of the intensity of spatial correlations for a fixed number of data samples. For small values of $\beta$ (and $\xi$) $W$ vanishes: the very weak spatial structure in the available data is not learned by the RBM. At large $\beta$, a place field emerges, focusing on a finite portion of the ring, and $W$ is non zero. The same transition is observed when $\beta$ is fixed and the number of training samples, $S$, is varied, see Fig.~\ref{ising1D_RBMtransition}(b). For few samples or, equivalently, large noise levels $r=N/S$, the RBM overfits the data and $W$ vanishes. For small values of $r$,  $W$ becomes non zero, signalling the emergence of  a place field focusing on a finite portion of the ring. 

This transition is an example of the very general mechanism of the so-called retarded learning phenomenon \cite{watkin}, also encountered in the context of random correlation matrices and the spiked covariance model. The connection with random matrices will be made explicit in Section \ref{secrlpt}.
	
\section{Learning data with multiple invariances}

\subsection{Data distribution: discretized XY model}

\begin{figure}[h]
  \begin{minipage}[c]{0.4\textwidth}
    \includegraphics[width=\textwidth]{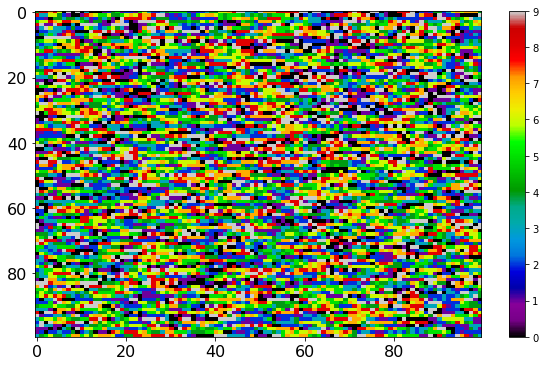}
  \end{minipage}\hfill
  \begin{minipage}[c]{0.5\textwidth}
    \caption{100 configurations (index along the $y$-axis) of the discretized XY model with $P=10$ Potts states over $N=100$ visible units (x-axis, index $i$ of the units) at inverse temperature $\beta=1.5$. Each colour refers to one of the 10 Potts states $v=0,1,...,9$.   } 
		\label{xy_config}
  \end{minipage}
\end{figure}

	The classical XY model is a popular model in statistical physics, used in particular to study topological phase transitions in two dimensions. We consider here the one-dimensional version of this model, which shows no such phase transition but is nonetheless very useful for our study due to the additional symmetry with respect to the Ising model. In the XY model each lattice site $i$ carries an angle $\theta_i\in [0,2\pi[$ with respect to some arbitrary, fixed direction. The energy function reads, up to a scale factor that can absorbed in the temperature definition,
	\begin{equation}
	E (\theta_1,\theta_2,...,\theta_N)= -\sum_{i=1}^{N}\cos(\theta_i - \theta_{i+1})
	\end{equation}
	with periodic boundary condition $\theta_{N+1} = \theta_1$. We then discretize the set of angle values in multiples of $2\pi/P$, where $P$ is an integer. The resulting model is a Potts model over the $N$ integer-valued variables $v_i=0,1,2,...,P-1$, with probability distribution (with periodic boundary conditions)
	\begin{equation}{\label{xy_pro}}
	p_{data}(v_1,v_2,....,v_N) = \frac{1}{\mathcal{Z}}e^{\, \beta\sum_{i=1}^{n}M(v_i,v_{i+1})}
	\end{equation}
	where the interaction kernel $M$ mimics the XY energy function,
	\begin{equation}
	M(v,v') = \cos\left(\frac{2\pi}{P}(v-v')\right) \ ,
	\end{equation}
	and the partition function normalizes the distribution $p$.
 This distribution enjoys two symmetries, compare to the single symmetry of the Ising model in (\ref{isingsym}): for any integers $K$ and $L$ we have,
 \begin{equation}\label{XYsym}
p_{data}(v_1,v_2,...,v_N) = p_{data}(v_{k+1}+L,v_{k+2}+L,...,v_{k+N}+L) \ ,
\end{equation} 
where $i+k$ and $v+L$ are to be intended, respectively, modulo $N$ and $P$. Figure~\ref{xy_config} shows a set of 100 configurations over $N=100$ sites, generated independently and at random from this model for $P=10$.

\subsection{Symmetry-breaking in both spaces}

\subsubsection{Case of a single hidden unit}

	We consider a RBM with $N=100$ visible Potts-type units $v_i$, which can take one out of $P=10$ values, and with $M=1$ hidden unit. The weights $w_{i,\mu=1}$ is now a vector $w_{i,1,v}$, with $i\in[1,2,...,100]$ and $v\in[1,2,...,10]$. The component $w_{i,1,v}$ of this vector is the connection between the hidden unit and the visible unit $i$ when it carries the Potts state $v$. 
	
\begin{figure}[h!]
		\centering
		\textbf{(a)}
		\includegraphics[width=0.45\linewidth,valign=t]{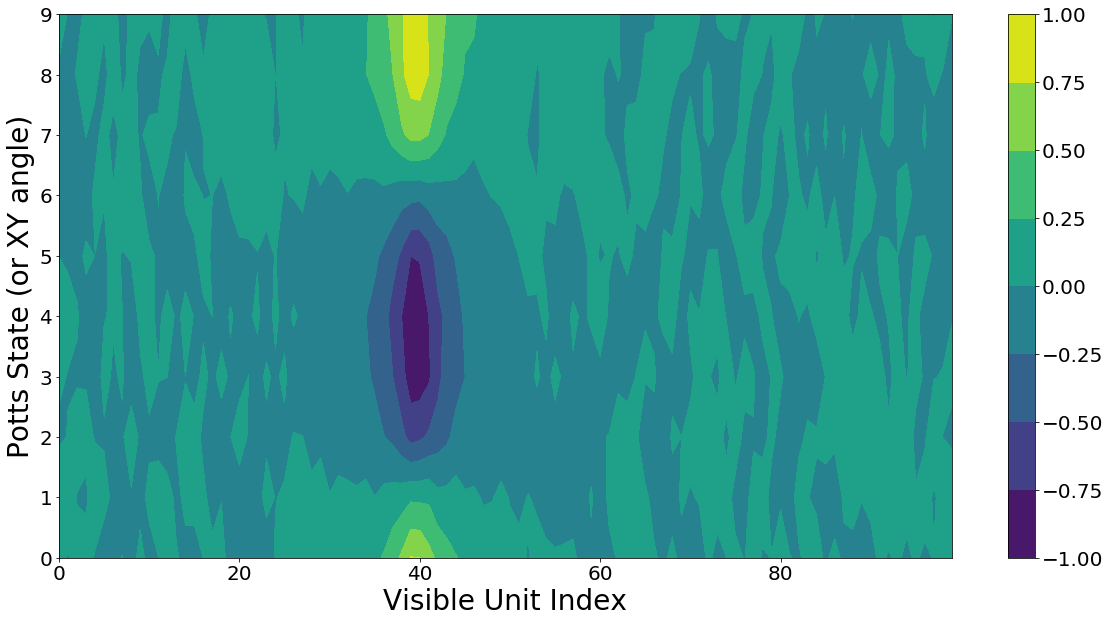} 
		\hskip .5cm
		\textbf{(b)}
		\includegraphics[width=0.45\linewidth,valign=t]{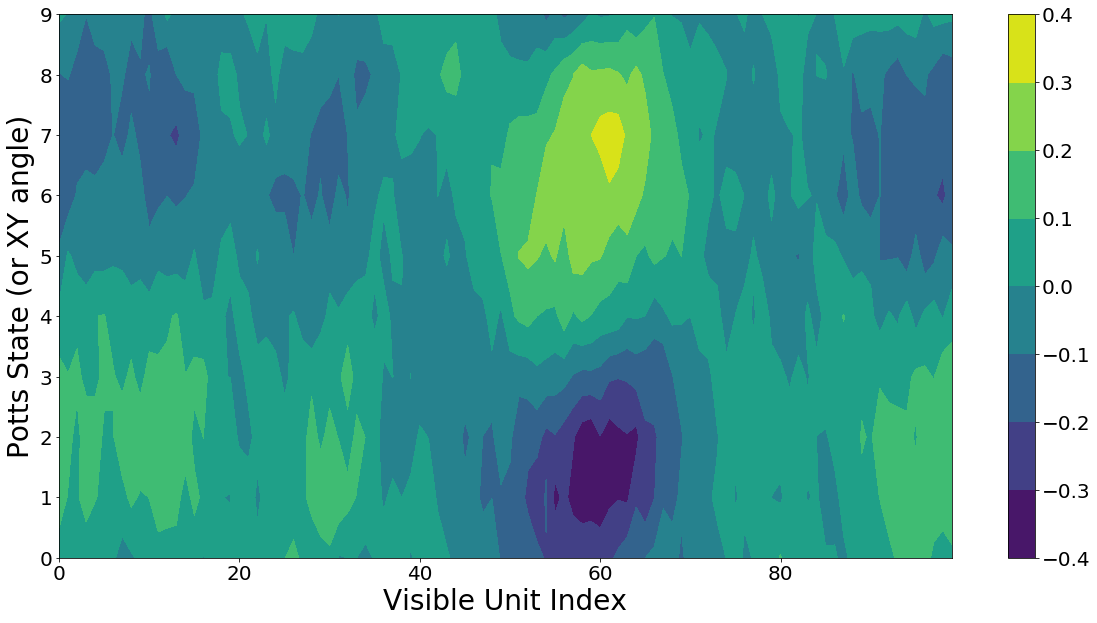}
		\textbf{(c)}
		\includegraphics[width=0.45\linewidth,valign=t]{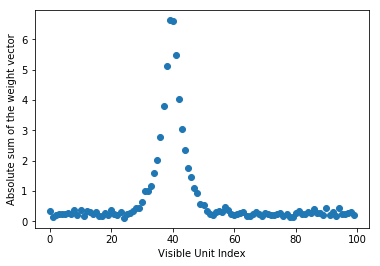} 
		\hskip .5cm
		\textbf{(d)}
		\includegraphics[width=0.45\linewidth,valign=t]{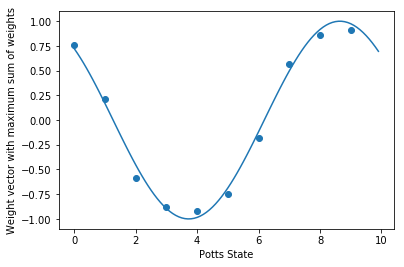}
		\textbf{(e)}
 		\includegraphics[width=0.97\linewidth,valign=t]{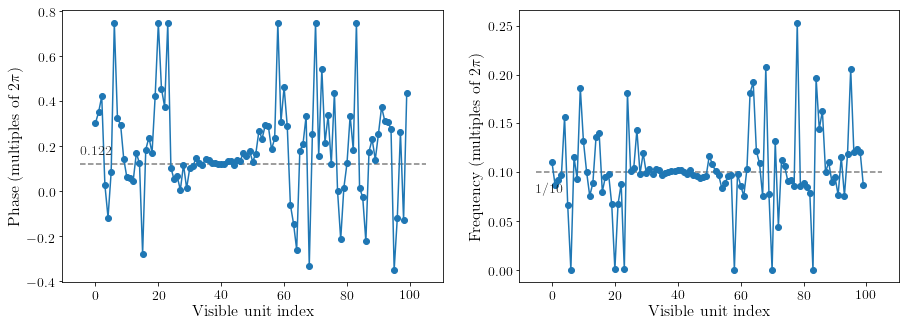} 	
		\caption{ {\bf (a).} Contour plot of the weights of the RBM with a single real-valued hidden unit with double-well potential trained on the XY model discretized by the Potts model. The $x$-axis shows space (index $i$ of the visible units), while the $y$-axis refers to angles (Potts state $v$). Parameters: $P= 10$ Potts states, $N=100$ visible units, $M=1$ hidden unit, trained on 100,000 configurations, learning rate $\nu=0.01$ and batch size $S=100$ trained over 100 epochs.  {\bf (b).} Same as {\bf (a)} for a binary hidden unit. 
		{\bf (c).} Angular modulation $W^{angular}_i$ of the weight vector fas a function of the space location $i$, see (\ref{op2}.  Same parameters as in Fig.~\ref{xy_RBM}(b) for the RBM with real-valued hidden unit. There is a clear strong space localization with a peak centered in unit $i=39$.    
		{\bf (d).} Weight vector  $w_{i=39,\mu=1,v}$ as a function of the angular-Potts variable $v$. The  line represents the cosine function with frequency = $2\pi/10$ as expected, with the best fit of the phase.  
		{\bf (e).} Phases $\varphi_i$ vs. site index $i$. Gray dotted line is the phase of the above fitting cosine. {\em (Right:)} Frequency $\omega_i$ vs. site index $i$. Gray dotted line is again the frequency of $2\pi/10$ of the cosine fit above, which is what one would expect from system with 10 Potts states. See text for the definition of the fitted frequencies and phases. The phase and the frequency is constant across the size of the receptive field, that is all the spins look in the same direction.}
	\label{xy_RBM}
	\end{figure}

	We first train a RBM with a single hidden unit $h_1$, which takes real values and is submitted to a double-well potential. Figure~\ref{xy_RBM}(a) shows the weights obtained after training from a very large number of configurations, starting from small white noise initial conditions for the $w_{i,1,v}$. We observe a strong modulation of the weights in the space and angle directions, achieving peak values around some site $i$ and angle $v$. Similar results were found for a binary-value hidden unit, $h_1=\pm 1$, with a slightly weaker localization of the weights and at a different location, see Fig.~\ref{xy_RBM}(b). In the following, we show results obtained for the RBM with the real-value hidden unit only.

	Since the interaction matrix $M$ in the Potts model takes the cosine function form, our RBM should learn the same functional dependence from the data samples. We show in  Fig.~\ref{xy_RBM}(c)  the quantity
	\begin{equation}\label{op2}
	W^{angular}_i=\left| \sum _{v=0}^{P-1} w_{i,1,v} \right| \ ,
	\end{equation}
	which measures the angular modulation of the weights on each site $i$. We see a strong space localization around $i=39$, because the weights only take non-zero values near that location. This location is arbitrary and similar to the place-field formation accompanying the breaking of translation symmetry  over space observed for the Ising model. In addition, at the location of the maxima, the weight vector is very well approximated by a cosine function, see Fig.~\ref{xy_RBM}(d). The RBM has learned the correct frequency equal to $2\pi/10$, and the phase takes an arbitrary value. Indeed, the phase in a free parameter due to the invariance against choices of $L$ in (\ref{XYsym}). 
	
To obtain a more precise picture of the receptive field, we then consider, for each site $i$, the $P$-dimensional vector of the weights $w_{i,\mu=1,v}$. We then fit this vector with a cosine function  of adjustable frequency and phase, referred to as, respectively, $\omega_i$ and $\varphi_i$.  We show, as functions of the site index $i$, the periods $\omega_i$ and the phases $\varphi_i$ in Fig.~\ref{xy_RBM}(e). We observe that the period takes the expected value $2\pi/P$ over the receptive field (sites ranging approximately between $i=30$ and 50). Similarly, the phase is constant (and takes an arbitrary value) over the same region of space. Informally speaking,  when the hidden unit is on, all the XY spins supported by the sites in the receptive field point to the same direction.

\subsubsection{Case of multiple hidden units}

\begin{figure}[h]
		\centering
		\textbf{(a)}
		\includegraphics[width=0.4\linewidth,valign=t]{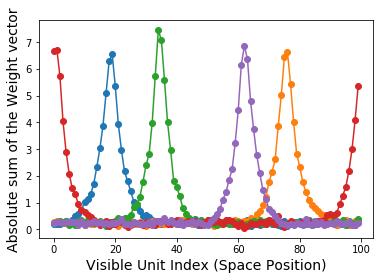} 
		\hskip .5cm
		\textbf{(b)}
		\includegraphics[width=0.42\linewidth,valign=t]{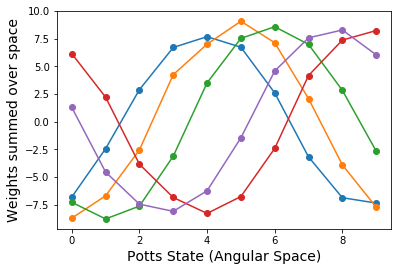} 		
		\caption{{\bf (a)} Angular modulation $W^{angular}_{\mu,i}$ (similar to (\ref{op2}), but the hidden-unit index $\mu$ runs from 1 to 5) vs. space location $i$. All the parameters are same as those of Fig.~\ref{xy_RBM} but with $M=5$ Real valued hidden unit with double well potential. There is a clear strong space localization and the receptive fields for the different units show mutual repulsion of their weight vectors.
			{\bf (b)} Weight vectors at the maximum visible unit index $i_{max}(\mu)$ for the respective hidden units $\mu$. The different curves each lie on a  cosine function with frequency = $2\pi/10$ as expected but with phases showing equal separations when ranked in increasing order. }
		\label{xy_multiple}
	\end{figure}

	We also train a RBM with $M=5$ Real valued hidden unit with double well potential, with results shown in Fig.~\ref{xy_multiple}. We see that the receptive fields of the hidden units are mutually separated in space, and show the same phenomenon of   repulsion between the units observed for the Ising data. In addition, the angular dependence of the five weight vectors exhibit the same frequency (equal to $2\pi/10$), but the phases show also a nice equi-separation due to repulsion along the angular direction. 
	
	Though we expected to see a diffusion of the receptive fields both along the spatial and angular dimensions for very large learning times, we did not observe this phenomenon even with RBM trained with 1,000,000 samples. This is likely due to the fact that the landscape is still rough for this amount of data, and diffusion remains activated. We have not tried to increase the number of samples because of the high computational cost.

\subsection{Differentiated retarded learning transitions} 

In this section, we show that RBM trained with data generated by the discretized XY model shows retarded learning phase transitions. However, as there are two potential symmetry breaking directions in this model, one corresponding to the angular space and the other to the positional space, the breaking of symmetry along these direction may take place at two different values of the noise ratio $r=N/S$, {\em i.e.} for different number of samples in the data set used for training. The reason is that the number of Potts states in the angular direction, $P$, may largely differ from the number of sites on the lattice, $N$.  Consequently, the effective system sizes along the two directions are different.

\begin{figure}[h]
		\centering
		\includegraphics[width=\linewidth]{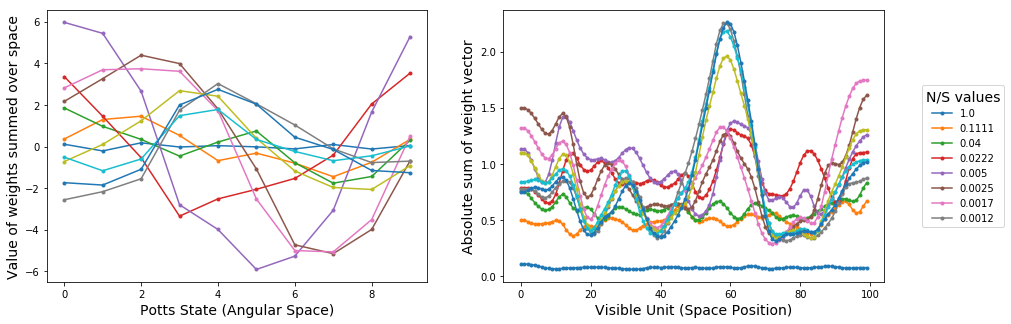}  
		\caption{The phenomenon of differentiated retarded learning phase transitions, for various noise ratios $r=N/S$, where $S$ is the number of training samples. 
		{\bf (a) Angular direction:} The y-axis is the sum of the components $w_{1,i,v}$ of the weight vector over all spatial location $i=1,...,N$. The x-axis shows the discretized angular states $v=0,1,...,P-1$.
		{\bf (b) Spatial direction:} The y-axis is the sum of the components $w_{1,i,v}$ of the weight vector over all angular variables $v=1,...,P$. The x-axis shows the lattice site $i=1,2,...,N$.	Parameters for data generation and sampling: $N=100$, $P=10$, $\beta=1.5$, $M=1$ (Bernoulli hidden unit), $\nu=0.1$, $S_{batch}=100$, trained for 20 epochs.}
		\label{xy_weights_transition}
	\end{figure}

This phenomenon of differentiated retarded learning phase transitions is reported in Fig.~\ref{xy_weights_transition}. We show in panel (a) of the figure the spatial modulation defined through,
	\begin{equation}\label{op3}
	W^{spatial}_v=\sum _{i=1}^{N} w_{i,1,v} \ ,
	\end{equation}
as a function of the Potts angular state variable $v$. We observe that for large $r$, the spatial modulation vanishes all over the angular space: low amount of data are not sufficient for the RBM to capture the angular correlations in the configurations. For large enough data set ($r<0.033$) the spatial modulation shows a clear dependence on $v$. We then show in panel (b) of  Fig.~\ref{xy_weights_transition} the angular modulation $W_i^{angular}$ as a function of the lattice site index $i$ for varied levels of sampling noise, $r$.  Again, for large $r$, no modulation is seen. However, for very small noise levels $r<0.002$, we do observe that $W_i^{angular}$ is peaked around some well defined site $i$. Interestingly, in the range $0.002 < r< 0.008$,  the angular modulation does not significantly vary over space, while the  spatial modulation varies over angles, compare panels (a) and (b). We conclude that, for intermediate ranges of values of $r$, the RBM has created a place-field along the angular direction, but not along the spatial direction.
	
	To test the generality of the phenomenon of differentiated transitions, we also generated data samples from variants of the discretized XY model. We modified the XY model in terms of changing the interaction matrix $M$ in (\ref{xy_pro}) from the cosine function to short range couplings, and also we changed the Hamiltonian to include not only nearest neighbor couplings but also long range couplings in the positional space. The resulting models display a variety of phase transitions in the RBM weights after training, with positional symmetry breaking arising before (for smaller amount of training data) angular ordering in some cases (not shown).

\section{Theoretical analysis}

Hereafter, we study analytically the dynamics of learning of the weights of the RBM with binary hidden units when trained with data. Two limit cases will be considered:
\begin{itemize}
\item The case of few data, which allows us to establish the connection with random matrix theory and  the so-called retarded learning transition;
\item The case of a large amount of data, with weak correlations, which we analyze in detail to understand the formation and shape of the place field, as well as the interactions between different place fields arising through learning.
\end{itemize}
While we will focus on the learning dynamics of the weights, we assume that the RBM has correctly learned the local fields, so we will set $b_i=c_\mu=0$ from the beginning in the case of unbiased binary data $v_i=\pm 1$. In addition, we assume that hidden units are of Bernoulli type, $h_\mu=\pm 1$. The log-likelihood therefore reads	
		\begin{equation}
		\log \mathcal{L} =\left\langle \sum_{\mu=1}^M \log  \, \cosh \left(\sum_{i=1}^N w_{i \mu } \,v_i \right) \right\rangle_{data} - \log {\cal Z}(\{w_{i\mu }\})\ ,
		\end{equation}
		where the partition function is
		\begin{equation}
		{\cal Z}(\{w_{i\mu }\}) =\sum _{\{ v_1,v_2,...,v_N\}} \prod_{\mu=1}^M  \cosh \left(\sum_{i=1}^N w_{i\mu } \,v_i \right)  \ .
		\end{equation}
Taking the partial derivative with respect to $w_{\mu i}$ we get the following expression for the gradient of the log-likelihood:
\begin{equation}\label{gradlldyna}
			\frac{\partial \log \mathcal{L} }{\partial w_{i\mu }} =\left\langle v_i \, \tanh \left(\sum_{j=1}^N w_{j\mu } \,v_j \right) \right\rangle_{data} - \frac 1{{\cal Z}(\{w_{i\mu }\}) }\, 
			\sum _{\{ v_1,v_2,...,v_N\}} v_i\,  \sinh \left(\sum_{j=1}^N w_{j\mu } \,v_j \right)   \prod_{\lambda (\ne \mu)} \cosh \left(\sum_{j=1}^N w_{i\lambda } \,v_j \right)   \ .
\end{equation}
The continuous-time dynamical equations for the evolution of the weights during training, assuming that the batch size is maximal, {\em i.e.} that all the data are used for training, are
\begin{equation}\label{dynaw}
			\frac{d w_{i \mu }}{dt} = \nu \; \frac{\partial \log \mathcal{L} }{\partial w_{i\mu }}  \ ,
\end{equation}
where $\nu$ is the learning rate.

\subsection{Few data: Small weight expansion and the retarded learning transition}\label{secrlpt}

\subsubsection{Linearized equations of  the dynamics} 

In this Section, we assume that the weights have initially very small (random) values. For small enough learning times, we may linearize the dynamical equations (\ref{dynaw}).  We obtain
 \begin{equation}\label{dynaw2}
			\frac{d w_{i\mu }}{dt} = \nu \; \bigg( \sum_{j=1}^N C_{ij} \, w_{j\mu }  -  w_{i\mu } \bigg)    \ ,
\end{equation}
where 
\begin{equation}\label{Cemp}
C_{ij} = \langle v_i \,v_j \rangle_{data} 
\end{equation}
is the empirical covariance matrix estimated from the data. Let $\Lambda$ be the largest eigenvalue of $C$, and $\bf e$ the associated eigenvector, with components $e_i$. As the diagonal elements $C_{ii}$ are equal to unity ($v_i^2=1$), we have that $\Lambda >1$, unless $C$ is the identity matrix and the data shows no correlation at all. Hence, according to (\ref{dynaw2}), all weight vectors ${\bf w}_{\mu} = \{ w_{1\mu },w_{2\mu }, ..., w_{N\mu }\}$ align along $\bf e$; this result holds within the linear approximation, and is therefore expected to be valid at short times only.

Let us consider the noise ratio $r=N/S$, equal to the number of visible units (system size) over the number of training samples. 
For bad sampling (large $r$),  the empirical covariance matrix can be approximated by the covariance matrix of a null model, in which all $N$ visible units are independent and unbiased: $v_i$ is equal to $\pm  1$ with equal probabilities ($=1/2$), independently of the other $v_j$'s. The asymptotic distribution of the eigenvalues of such a random matrix has a special form, called the Marcenko Pastur (MP) spectrum, whose right edge (top eigenvalue) is given by 
\begin{equation}\label{MP}
	\Lambda_{noise} = \Lambda_{MP} = \left(1+\sqrt{r}\right)^2
\end{equation}
and the corresponding top eigenvector $\bf e$ has random, Gaussian distributed components. 

Conversely, for good sampling (small $r$), we expect the empirical covariance to be similar to the covariance matrix computed from the model distribution $p$ from which data were generated. Due to the translational invariance of $p$, its top eigenvector ${\bf e}_{Model}$ has the same symmetry: ${\bf e}_{model}=(1,1,...,1)$, up to a normalization factor. Hence, we expect $\bf e$ to be similar to ${\bf e}_{Model}$ and be roughly uniform. In the double, large $N$ and large $S$ limit, the two regimes may be separated by a sharp transition, taking place at a critical value of $r$. To locate this value, we compute below the top eigenvalue of the model covariance matrix, and compare it to its MP counterpart (\ref{MP}). The crossover between the bad and good sampling regimes takes place when both eigenvalues are equal.

\subsubsection{Case of Ising data}

Let us consider the case of the one-dimensional Ising model. When a large number of configurations is available, we have $C_{ij}=(\tanh \beta)^{|i-j|}$, see (\ref{isinghj}). Due to the rotational invariance,  the top-eigenvector, ${\bf e}_{Ising}$ has all its components equal. Therefore the top eigenvalue of the covariance matrix is
	\begin{equation}
	\Lambda _{Ising}(\beta)= \sum_{j=1}^{N}C_{ij}= 1 + 2\,\big[ \tanh \beta +  \tanh ^2 \beta  +  \tanh ^3 \beta+  ...	\big]\approx \frac{1+\tanh\beta}{1-\tanh\beta}= e^{2\,\beta} \ .
	\end{equation}
When the inverse temperature $\beta$  is small, this `signal' eigenvalue is smaller than  the `noise' eigenvalue $\Lambda _{MP}$ (\ref{MP}), locating the right edge of the MP spectrum. In this case, we expect the top eigenvector $\bf e$ of the empirical covariance matrix $C$ to be noisy, and not to capture the correlation between the Ising variables $v_i$. In this regime, the RBM overfits the data and no receptive field with a localized weight structure can emerge. As $\beta$ increases above
\begin{equation}\label{br}
\beta (r) = \log  \left(1+\sqrt{r}\right) \ ,
\end{equation}
the signal eigenvalue $\Lambda _{Ising}(\beta)$ becomes larger than the MP edge, and we expect the top eigenvector of $C$ to have comparable component and be similar to ${\bf e}_{Ising}$.

\begin{figure}[h]
		\centering
		\textbf{(a)}
		\includegraphics[width=0.45\linewidth,valign=t]{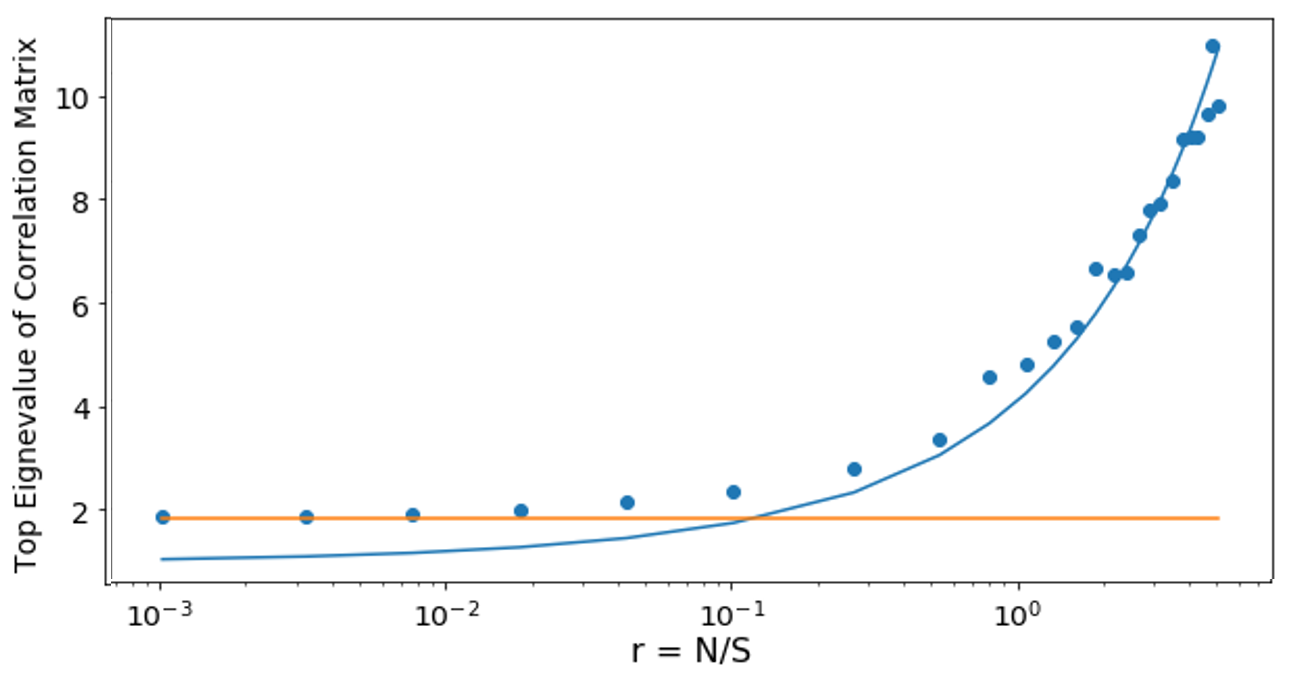} 
		\hskip .5cm	
		\textbf{(b)}
		\includegraphics[width=0.45\linewidth,valign=t]{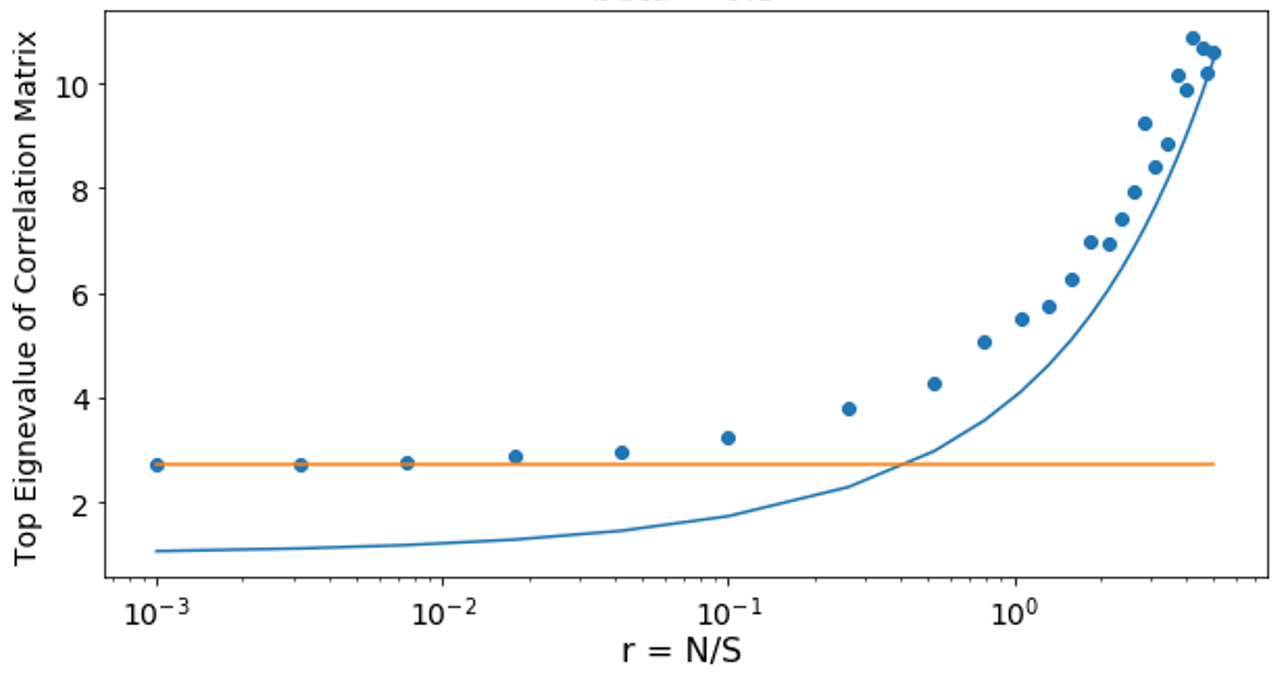}
		\textbf{(c)}
		\includegraphics[width=0.5\linewidth,valign=t]{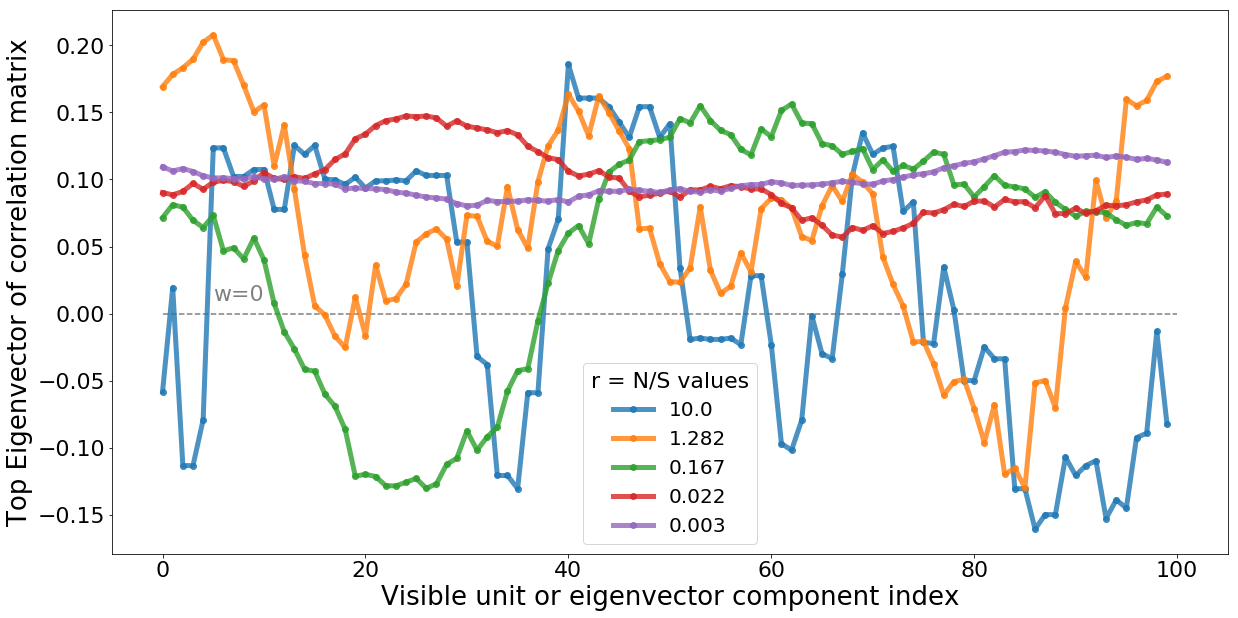} 
		\caption{{\bf (a)} Top Eigenvalue of the correlation matrix $C$ of data generated from the one-dimensional Ising model at inverse temperature $\beta=0.3$, as a function of the noise ratio $r$. The orange straight line is the top eigenvalue $\Lambda_{Ising}$,  corresponding to a perfectly sampled (infinite $S$) Ising model. The blue curve is the top eigenvalue $\Lambda_{noise}$ of the correlation matrix of the null model with independent variables.
		{\bf (b)} Same as panel (a) but for $\beta=0.5$.
		{\bf (c)} Top eigenvector of the correlation matrix $C$ (\ref{Cemp}) of the configurations of the one-dimensional Ising model at fix $\beta$ but with different numbers $S$ of samples. Ising model samples to calculate the correlation matrix were generated at $\beta=1$ for $N=100$ spins.}
		\label{ising1DtopEVal}
	\end{figure}

The above statement is corroborated by the results shown in Fig.~\ref{ising1DtopEVal}(Top), which shows the top eigenvalue $\Lambda$ of the correlation matrix $C$ (\ref{Cemp}) as a function of the noise ratio, $r=N/S$, where $S$ is the number of samples. For large $r$ (few samples), $\Lambda$ is very well approximated by $\Lambda_{noise}$, while, for small $r$ (many samples), $\Lambda$ gets very close to 
$\Lambda _{Ising}(\beta)$ as expected. The crossover between these two regimes takes place at values of $r$ such that $\beta\simeq \beta(r)$ (\ref{br}).

Figure~\ref{ising1DtopEVal}(c) shows how the top eigenvector of the data correlation matrix changes as more and more samples are considered. One clearly sees a phase transition  from a random vector to the uniform eigenvector ${\bf e}_{Ising}$.

\subsubsection{Case of XY data}
\begin{figure}[h]
\includegraphics[width=1.0\linewidth]{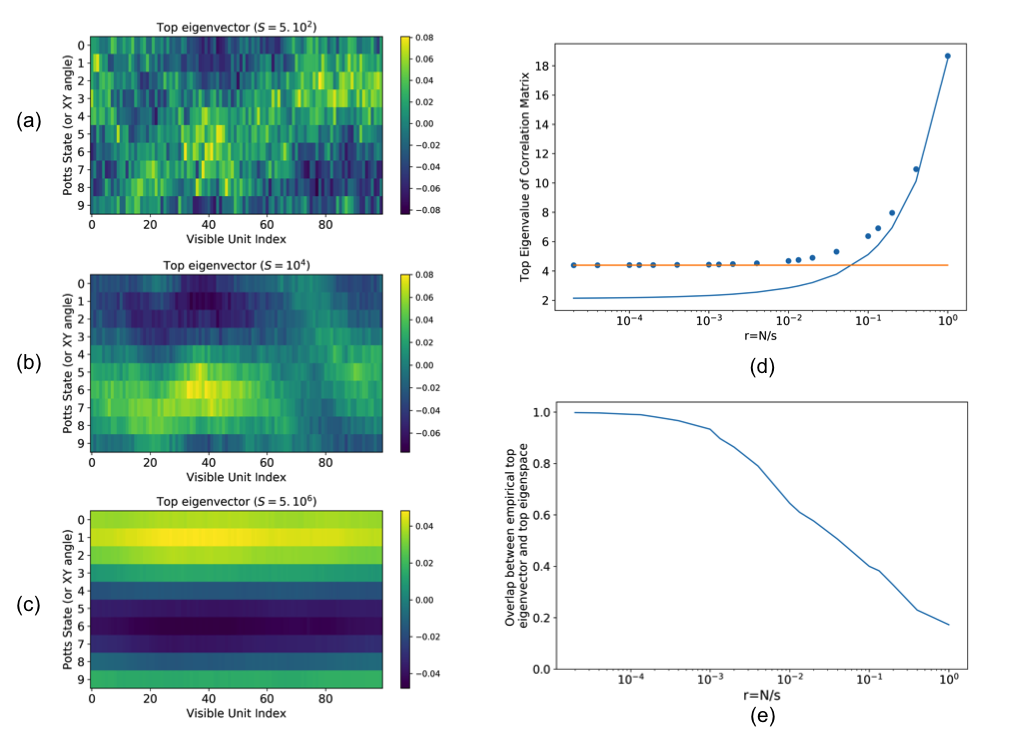} 
		\caption{{\bf (a),(b),(c)} Empirical top eigenvector of the correlation matrix $C$ of data generated from the XY model at inverse temperature $\beta=1.5$, for sample size $S=5\, 10^2$ {\bf (a)}, $S=10^4$ {\bf (b)}, $S=5 \, 10^6$ {\bf (c)}. {\bf (d)}. Corresponding top eigenvalue (dots) as a function of the sample size $r=\frac{N}{S}$. The orange straight line is the top eigenvalue $\Lambda_{XY}(\beta=1.5)$, corresponding to a perfectly sampled (infinite $S$) XY model. The blue curve is the top eigenvalue $\Lambda_{noise}$ of the correlation matrix of the null model with independent variables. The eigenvalues cross $r \simeq 0.066$. {\bf (e)}. Norm of the projection of the empirical top eigenvector $\hat{{\bf e}}$ in the space spanned by the perfect-sampling top eigenvectors ${\bf e^1}$,${\bf e^2}$,  $\sqrt{ (\hat{{\bf e}} . {\bf e^1 })^2+ (\hat{{\bf e}} . {\bf e^2 })^2 }$ }
\label{XYtopEVal}
\end{figure}

For the discrete XY model, the correlation matrix in the $r \rightarrow 0$ limit can be computed as well using the transfer matrix formalism. We find
\begin{equation}
C_{ij}(v,v') = \frac{1}{P-1} \sum_{p=1}^{P-1} \left(\frac{\lambda_p(\beta)}{\lambda_0(\beta)}\right)^{|j-i|} \cos \left(\frac{2 \pi p (v-v')}{P}\right) \ ,
\end{equation}
where
\begin{equation}
\lambda_p (\beta)= \sum_{v=0}^{P-1} \exp \left[ \beta \cos \left(\frac{2 \pi v}{P} \right)\right] \cos\left( \frac{2 \pi v p}{P}\right)\ .
\end{equation}
$C$ enjoys translational invariance along both axis, hence its eigenvectors are discrete 2D Fourier modes; after computation, we find that the top eigenvalue is
\begin{equation}
\Lambda_{XY} (\beta)= \frac{P}{P-1} \frac{\lambda_0(\beta)+\lambda_1(\beta)}{\lambda_0(\beta)-\lambda_1(\beta)}\ .
\end{equation}
with a corresponding eigenspace of dimension 2, spanned by $e^1_i(v) = \sqrt{\frac{2}{NP} }\cos \left( \frac{2 \pi v}{P} \right)$, $e^2_i(v) =\sqrt{ \frac{2}{NP}} \sin \left( \frac{2 \pi v}{P} \right)$. The top eigenvector is uniform over space, as for the Ising model, but not over the angular variables, see Fig.~\ref{XYtopEVal}(c).

The 'noise' eigenvalue is similarly given by the MP spectrum, although slightly modified: the  dimension to sample size ratio is now $\frac{P N}{S} = P r$, and in the $S \rightarrow \infty$ limit, the correlation matrix has top eigenvalue $\frac{P}{P-1}$ owing to the anticorrelations between Potts variables on the same site, $C_{i,i}(a,b) = -\frac{1}{P-1}, \; \; \forall \, a \neq b$. We obtain:

\begin{equation}
\Lambda_{noise} = \frac{P}{P-1} + \left( 1 + \sqrt{rP} \right)^2
\end{equation}
Similarly to the case of Ising data, when $\beta$ is small, the signal 'eigenvalue' is small compared to the 'noise' eigenvalue, and the empirical top eigenvector has a small projection in the space spanned by ${\bf e^1}$,${\bf e^2}$, see Fig.~\ref{XYtopEVal}(b,c,d). The crossover between the two regimes takes place at values of $r$ such that $\Lambda_{MP} \simeq \Lambda_{XY}(\beta)$. The first retarded learning transition of the RBM occurs in the same range of $r$, see Fig.\ref{xy_weights_transition}.

\subsection{Many data: Small  $\beta$ expansion}\label{betazero}

		After some training time, linear equation (\ref{dynaw2}) for the weights breaks down and non linearities must be taken into account \cite{decelle}. We derive below an approximation to the RBM dynamic learning equation (with $M=1$ or 2 hidden units) for the one-dimensional Ising models, which is exact for small (but non vanishing) inverse temperature $\beta$.  We show that this equation is free of any external parameters after appropriate rescaling of the weights. We compare the numerical solutions to this equation with the result of the training with RBM to find a parameter independent agreement with the shape and the structure of the weights. We also cast the equation into a continuous form, and formulate the system in terms of a standard Reaction-Diffusion instability problem with the weights as an inducer and the sum of weights squared as the repressor.

		\subsubsection{One Hidden Unit System: formation of receptive field}
		
\begin{figure}[h]
			\centering
			\textbf{(a)}
			\includegraphics[width=0.265\linewidth, valign=t]{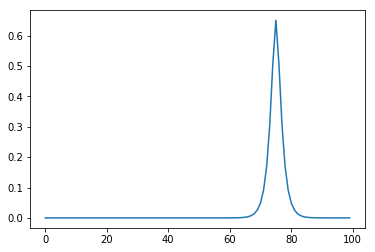} 
			\hskip .2cm
			\textbf{(b)}
			\includegraphics[width=0.28\linewidth, valign=t]{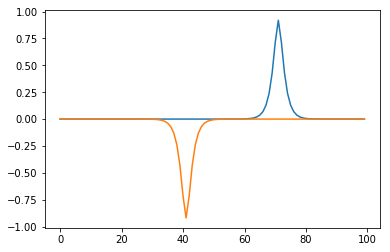}
			\hskip .2cm
			\textbf{(c)}
			\includegraphics[width=0.35\linewidth, valign=t]{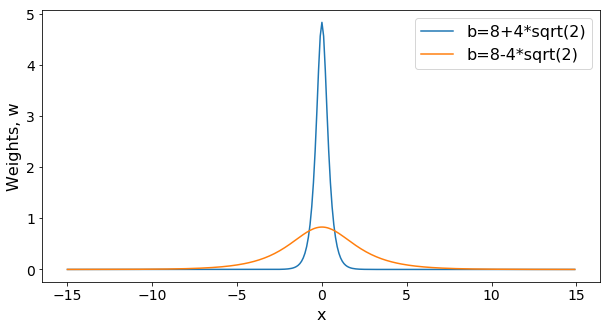}
			\caption{{\bf (a)} Stationary solution of the small $\beta$ equations (\ref{eq1}), describing the evolution of the weights $w_i(t)$ of a RBM with $M=1$ hidden unit trained over many configurations of the one-dimensional Ising model. There are $N= 100$ visible units. The results shown were obtained with 500 integration steps, starting from small amplitude white noise initial conditions for the weights.
			{\bf (b)} Same as (a), but for a RBM with $M=2$ hidden units. The two colors shows the weights corresponding to the two units.
			While the two peaks should be in principle diametrically opposed, i.e. at distance 50 from each other, their mutual repulsion is short ranged; in practice deviations from stationarity smaller than the numerical accuracy cannot be detected.
			{\bf (c)} Profiles of the stationary weight vector for a RBM with a unique hidden unit trained on data extracted from the one-dimensional Ising model at small inverse temperature $\beta$, see text. The two curves corresponds to the two candidate values for $b$. The solutions $b=b_+$ and $b=b_-$ are, respectively, unstable and stable against small fluctuations of the weights.}
			\label{numerical_soln1D}
	\end{figure}

For one hidden unit, equations (\ref{gradlldyna},\ref{dynaw}) become, after some elementary manipulation, 
		\begin{equation}\label{eq0}
		\frac{\partial \log \mathcal{L}}{\partial w_j} = \left\langle v_i \, \tanh \left(\sum_{j=1}^N w_j \,v_j \right) \right\rangle_{data} - \tanh w_i \ ,
		\end{equation}
where we have dropped the $\mu=1$ index for the sake of clarity. Expanding the hyperbolic tangents to the third powers of their arguments, we obtain
		\begin{equation}\label{eq00}
		\frac{\partial \log \mathcal{L}}{\partial w_j} =  \sum_j \langle v_i \, v_j \rangle_{data} \; w_j- \frac{1}{3}\sum_{j,k,l}\langle v_i\,v_j\,v_k\ v_l \rangle _{data}\; w_j\,w_k\,w_l-w_i+\frac{1}{3}w_i^3 + O(w^4)\ .
		\end{equation}
Let us now assume that a large number of samples is available. At the lowest order in $\beta$, we have
		\begin{equation}
		\langle v_i\, v_j\rangle = \left\{ \begin{array}{c c c} 1 & \text{if} & i=j\ , \\
		\beta & \text{if} & i = j\pm1 \ ,\\
		0 & & \text{otherwise}\ .
		\end{array}\right.
		\end{equation}
		and
		\begin{equation}
		\langle v_i\,v_j\,v_k\, v_l \rangle =  \left\{ \begin{array}{c c c} 1 & \text{if} & i=j, k=l\  \text{or any permutation}\ , \\
	\beta & \text{if} & i=j\pm 1, k=l\  \text{or any permutation}\ , \\
	\beta & \text{if} & i=j, k=l\pm 1\  \text{or any permutation}\ , \\
	0 & & \text{otherwise}\ .
		\end{array}\right.
		\end{equation} 
for, respectively, the 2- and 4-point correlations. we therefore obtain
		\begin{equation}
\frac{\partial \log \mathcal{L}}{\partial w_j} = \beta(w_{i+1} + w_{i-1}) - w_i\sum_k w_k^2 +w_i^3 +O( w^4, \beta\, w^3) \ .
		\end{equation}
Upon appropriate rescaling of the weights, $w_i\to w_i / \sqrt \beta$, and of the learning rate, $\nu \to \nu /\beta$, we obtain, in the small $\beta$ regime, the non trivial, parameter-free dynamical equation
		\begin{equation}\label{eq1}
		\frac 1\nu\frac{dw_i}{dt} = 	w_{i+1} + w_{i-1} - w_i\sum_k w_k^2 +w_i^3	\ .
		\end{equation}
The stationary solution of this equation is shown in Fig.~\ref{numerical_soln1D}(a). 	

This equation can be cast in a continuous form over space, where we use the Laplacian to describe spatial diffusion. The corresponding continuous partial differential equation reads
\begin{equation}\label{eq2}
		\frac 1\nu \frac{\partial w}{\partial t} (x,t) = \frac{\partial^2 w}{\partial x^2} (x,t) + \Big( 2 - b(t) \Big)\,w(x,t) +  w(x,t)^3\ ,
\end{equation}
where 
\begin{equation}\label{defbb}
		b(t) = \int_{0}^{L} w(x,t)^2\, dx \ .
\end{equation}
These coupled dynamical equations lead to a non-trivial spatial formation through the so-called {\em Turing Reaction Diffusion instability} mechanism \cite{book1}. The field $w(x,t)$ diffuses over space and activates itself (self-promoting, through the cubic term), but is inhibited by another species, $b$. This repressor  is diffusing with an infinite diffusion coefficient, {\em i.e.} is spatially uniform, and depends on $w$ through (\ref{defbb}). As $w$ grows due to self-activation, so does the repressor $b$, until $w$ reaches a stationary profile. We show in Appendix \ref{secapp} that the above dynamical equation satisfy the  general criteria for stable pattern formation.

\subsubsection{One hidden unit: Profile of the receptive field}

Consider the stationary continuous equation satisfied by the weights after learning, see (\ref{eq2}),	
\begin{equation}\label{eq22}
0=  \frac{d^2 w}{dx^2} (x) + \Big( 2 - b \Big)\,w(x) + w(x)^3\ .
\end{equation}
Multiplying by $\frac{d w}{dx}$ on both sides and integrating over $x$ we obtain that
	\begin{equation}
		E (x) \equiv \frac{1}{2}\left(\frac{d w}{d x}\right)^2 + w(x)^2 +\frac{1}{4}w(x)^4 -\frac{b}{2}\, w(x)^2 
	\end{equation}
has a uniform value $E_0$, independent of $x$. When $x \rightarrow \pm \infty$, both $w(x)$ and $\frac{dw}{dx}(x)$ tend to 0, which sets $E_0=0$. We deduce that	
	\begin{equation}\label{eq3}
 \frac{dw}{dx}(x) = \pm w(x) \sqrt{b-2 - \frac{w(x)^2}{2}} \ .
	\end{equation}
We now explicitly break the symmetry by fixing the centre of the peak of the weights in $x=0$, with $w(0)>0$,  $\frac{dw}{dx} >0$ for $x<0$, and $\frac{dw}{dx} <0$ for $x>0$. 
Imposing that the derivative of the weight with  respect to $x$ vanishes at its maximum, {\em i.e.} that $w$ is twice differentiable in $x=0$ gives 	
	\begin{equation}\label{eq4}
w (0)= \sqrt{2(b-2)} \ .
	\end{equation}
Integrating (\ref{eq3}) with condition (\ref{eq4}), we find
\begin{equation}{\label{defweight}}
	w (x) = \frac{\sqrt{2(b-2)}}{ \cosh \big( x \sqrt{b-2}\big)}\ .
\end{equation}
Using definition (\ref{defbb}) for $b$ we then find
\begin{equation}
b= \int_{-\infty}^{\infty}w(x) ^2\, dx = 4\sqrt{b-2} \ ,
\end{equation}
whose solutions are 	$b _\pm = 8 \pm 4\sqrt{2}$.  The corresponding profiles of the weights are shown in Fig.~\ref{numerical_soln1D}(c). We now study the stability of the solution under the time-dependent perturbation $w(x) \to w(x) + \epsilon(x,t)$, where $w(x)$ is given by (\ref{defweight}). According to equations (\ref{eq2})
and (\ref{eq22}), we have
\begin{equation}\label{eq23}
		\frac 1\nu \frac{\partial \epsilon}{\partial t} (x,t) = \frac{\partial^2 \epsilon}{\partial x^2} (x,t) + \Big( 2 - b \Big)\,\epsilon(x,t) -2\,  \bigg( \int dy \,w(y)\,\epsilon(y,t) \bigg)\, \epsilon(x,t)+  3\,w(x)^2\,\epsilon(x,t) \ .
\end{equation}
Multiplying by $w(x)$ and integrating over $x$, we get the following equation
\begin{equation}\label{eq24}
		\frac 1\nu \frac{d}{dt} \int dx\, w(x)\, \epsilon (x,t) = - \sqrt{b-2}\, \int dx\, w(x)\, \epsilon (x,t) \left( 8 - \frac {b}{   \cosh \big( x \sqrt{b-2}\big)^2}  \right)  \ .
\end{equation}
We deduce that the weight profile is stable if and only if $b<8$. Therefore, the $b=b_+$ solution is unstable against small variations of the peak amplitude near $x=0$, and the solution $b=b_-$ is the correct, stable one. Notice that the width of the peak of the weight, in the $\beta \rightarrow 0$ limit is finite according to expression (\ref{defweight}). This phenomenon was also observed by the RBM training results in Fig~\ref{rbm1Dising}(b), where the peak width obtained by linear fit (coefficient $b$) was also positive and finite.

\subsubsection{Two hidden units: interaction between receptive fields}

For two hidden units, equations (\ref{gradlldyna},\ref{dynaw}) become, after some simple manipulation, 
\begin{equation}\label{eq6}
\frac{\partial \log \mathcal{L}}{\partial w_{j1}} = \left\langle v_i \, \tanh \left(\sum_{j=1}^N w_{j1} \,v_j \right) \right\rangle_{data} - \frac{\tanh (w_{i1}+w_{i2})}{\displaystyle{ 1+ \prod _{k=1}^N \frac{\cosh (w_{k1}-w_{k2})}{ \cosh (w_{k2}+w_{k2})}}} - \frac{\tanh (w_{i1}-w_{2i})}{\displaystyle{ 1+ \prod _{k=1}^N \frac{\cosh (w_{k1}+w_{k2})}{ \cosh (w_{k1}-w_{k2})}}} \ ,
\end{equation}
together with a similar equation for the weight vector $\mu=2$ obtained by swapping the hidden-unit indices $1$ and $2$. Notice that this equation simplifies to (\ref{eq0}) when the weight vector $\mu=2$ is set to zero,  {\em i.e.} when $w_{i2}=0$ for all visible units $i$, and the number of hidden units is effectively $M=1$.

Let us now expand (\ref{eq6}) in powers of the weights. The first term on the right hand side of the equation (involving the average over the data distribution) has the same expansion as in the $M=1$ case above, see (\ref{eq00}). For the second term, using
\begin{equation}
1+ \prod _{k=1}^N \frac{\cosh (w_{k1}-w_{k2})}{ \cosh (w_{k1}+w_{k2})} = 2 -2 \sum _{k=1}^N w_{k1}\, w_{k2} + O(w^4)
\ ,
\end{equation}
and rescaling the weights, $w\to w / \sqrt \beta$, and the learning rate, $\nu \to \nu /\beta$, as before, we obtain
\begin{equation}\label{eq7}
\frac 1\nu\frac{dw_{i1}}{dt} = w_{i+1,1} + w_{i-1,1} - w_{i1}\, \sum _{k} w_{k1}^2+w_{i1}^3 + w_{i1}\, w_{2i}^2  - w_{i2} \, \sum _{k} w_{k1}\, w_{k2}  \ .
\end{equation}
Similarly, we find
\begin{equation}\label{eq7b}
\frac 1\nu\frac{dw_{i2}}{dt} = w_{i+1,2} + w_{i-1,2} - w_{i2}\, \sum _{k} w_{k2}^2+w_{i2}^3 + w_{i2}\, w_{i1}^2  - w_{i1} \, \sum _{k} w_{k1}\, w_{k1}  \ .
\end{equation}
The last two terms in the two equations above encode the couplings between the weight vectors attached to the two units. The stationary solutions of these equations are shown in Fig.~\ref{numerical_soln1D}(b). In practice, we observe that the numerical solutions for the weight profiles attached to the two units can have
any relative separation between their peaks as long as it is larger than approximately one peak width. The repulsion between the
peaks is indeed short range, hence the convergence to the diametrically opposed configuration is
really slow.

These equations can be turned into two partial differential equation over the continuous space continuous where we resort to the Laplacian to describe spatial diffusion:
\begin{eqnarray}\label{eq20}
		\frac 1\nu \frac{\partial w_1}{\partial t} (x,t) &=& \frac{\partial^2 w_1}{\partial x^2} (x,t) + \Big( 2 - b_1(t) \Big)\,w_1(x,t) -c\, w_2(x,t) +  w_1(x,t)\Big( w_1(x,t)^2+w_2(x,t)^2\Big) \ , \nonumber \\
		\frac 1\nu \frac{\partial w_2}{\partial t} (x,t) &=& \frac{\partial^2 w_2}{\partial x^2} (x,t) + \Big( 2 - b_2(t) \Big)\,w_2(x,t) -c\, w_1(x,t) +  w_2(x,t)\Big( w_1(x,t)^2+w_2(x,t)^2\Big) \ , 
\end{eqnarray}
where 
\begin{equation}\label{defbb2}
		b_1(t) = \int_{0}^{L} w_1(x,t)^2\, dx \ , \quad b_2(t) = \int_{0}^{L} w_2(x,t)^2\, dx \ , \quad c(t) = \int_{0}^{L} w_1(x,t)\, w_2(x,t)\, dx \ ,
\end{equation}
This system describes two diffusing species $w_1$ and $w_2$ which are, respectively, self-inhibited by $b_1$ and $b_2$, while coefficient $c$ corresponds to cross-inhibition. The diffusion coefficient for $b_1, b_2, c$ tend to infinity and their concentrations are uniform in space. As in the single species (single hidden-unit) case, this dynamical system leads to the stable production of a non trivial pattern over space, corresponding to the emergence of two place fields, see Appendix \ref{secapp}.

\section{Conclusion}

In this work, we have studied the unsupervised learning of simple data distributions, enjoying one or two continuous symmetries, with a RBM. Contrary to standard approaches in machine learning, {\em e.g.} convolutional networks, we have not tried to factor out, and hardwire these symmetries in the network. On the contrary, our objective was to see how the symmetries affected the representations of the data and were effectively learned by the machine. This approach is motivated by the fact that most invariances in complex data are actually unknown and it is important to understand how well they can be captured in practice.

In the case of a single hidden (latent) variable, our main observation is that learning is accompanied by a symmetry breaking in the weight space. The hidden unit concentrates only on a small portion of the data configurations; the size of this receptive field is the length over which the variables in the data configurations are correlated. The symmetry is dynamically restored at long times through the diffusion of the receptive field, allowing it to span the whole data manifold. This phenomenon is strongly reminiscent of the concept of continuous attractor (CA) in the context of recurrent neural networks in computational neuroscience \cite{amari,tsodyks,sompo}, with the major differences that (1) CA usually refer to low-dimensional attractors in the (high-dimensional) space of neural activities, while the CA emerging here defines a low-dimensional manifold in the {\em weight space}, and (2) accordingly, the dynamics considered is the {\em learning} dynamics acting on weights and not the usual neural dynamics modifying activities. In the case of multiple hidden units, the CAs attached to these units are locked in: weight bumps diffuse coherently along their CA's (Fig.~\ref{smooth1D_3u}), maintaining their relative phases due to mutual repulsive interactions. The resulting multi-unit CA has therefore the same (low) dimension as the underlying symmetry in the data. In practice, however, repulsion is short ranged and may effectively lead to partial decoupling, see Fig.~\ref{numerical_soln1D}(b), and to an increase in the CA dimension. If the number of hidden units is sufficiently large (of the order of the number of visible units over the correlation length) the RBM hidden configurations are effective, coarsegrained version of the data configurations.

An important condition for this CA in the weight space to emerge is that the number of available data exceeds some critical value depending on the configuration size and the intensity of their intrinsic correlations. This phenomenon is a manifestation of the general mechanism of the so-called retarded-learning phase transitions \cite{watkin}, in which a symmetry-breaking direction (here, in the weight space) is inferred when the ratio of the number of data and of the system size is larger than some signal-dependent (here, the level of intrinsic correlation in the data) threshold. Interestingly, in the presence of multiple invariances, the thresholds associated to these symmetries need not coincide. In such situations, the receptive field will be localized along one dimension in the input space and extended along the other, as seen for the XY model in this work.

It is tempting to make an analogy with recent experimental results on three-dimensional encoding by place cells \cite{treves}. When a rodent explores a set of horizontal ($x$ direction) and ($z$ direction) vertical planes, place cells emerge with place fields localized in either or both planes \cite{Casali19}. Yet, if the motion along the $x$ and $z$ axis is not independent, localization can be lost along one of the two directions. For instance, when motion takes place along a helicoidal ramp ($x$ being the angle  in the plane perpendicular to the helix axis $z$), place cells seem to be localized in the angular space and much less so along the vertical axis \cite{Hayman11}. Due to the geometry of the helix, it is reasonable to assume that inputs related to path integration as well as to visual flow are strongly correlated for similar angles (corresponding to a small displacement on the ramp) and much less correlated for small translation along the $z$-axis, which requires a large physical displacement. It would be interesting to see what happens if the ramp axis is tilted and not vertical any longer. Based on the analogy with the differential retarded-learning transitions, one would expect that place fields are columnar along the ramp axis, and become therefore localized (albeit with different areas) along both $x$ and $z$ directions.

While the analogy with place cells and symmetry-broken hidden units is tempting, establishing a solid connection between our results and neuroscience is far from obvious. Though place cells are known to rely, for their establishment, on various sources of sensory information (including visual and path-integration inputs \cite{vrokeefe}), the mechanisms  underlying the corresponding unsupervised learning processes are far from being elucidated. It is, from this point of view, remarkable that various unsupervised learning rules \cite{sengupta,benna,gerstner} agree with the two main features emerging from the Maximum Likelihood (ML) procedure for RBM studied here, namely the existence of (1) localized receptive field focusing on a subsets of strongly correlated inputs, and of (2) cross-inhibition between hidden units during the learning phase, which makes their place/receptive fields repell each other and forces them to cover as much as possible the input space (Fig.~\ref{rbm1Dising3units}). Achieving a more precise understanding of how general this scenario is, and how it extends to deeper architectures {\em i.e} with more neural layers would be very interesting.

\vskip .3cm
\noindent 
{\bf Acknowledgments.} This work was partly funded by the ANR project RBMPro CE30-0021-01 and the HFSP project RGP0057-2016. M.H. benefited from a fellowship from the ICFP Labex of Department of Physics at ENS. J.T. acknowledges partial support by a fellowship from the Edmond J Safra Center for Bioinformatics at Tel Aviv University.

\appendix

\section{Conditions for pattern formation}
\label{secapp}

The continuous partial differential equation (\ref{eq2}) along with (\ref{defbb}) describes the evolution of the field $w(x,t)$ in space and time. This equation will only lead to a non-trivial stable steady state pattern if certain conditions that we make explicit below are fulfilled.

We start from  (\ref{defbb}) and differentiate this equation with respect to time (setting $\nu=1$ to lighten notations) to get
\begin{equation}
\frac{db(t)}{dt} = 4\,b(t) - 2\,b(t)^2 + 2\int_{0}^{L} w(x,t) \frac{\partial ^2}{\partial x^2} w(x,t)\,dx + 2\int_{0}^{L}w^4(x,t)\,dx \ .
\end{equation}
The non-trivial, uniform fixed point this equation is
\begin{equation}
w^* = \sqrt{\frac{2}{L-1}} \quad \text{and} \quad b^* = \frac{2L}{L-1} \ .
\end{equation} 
One can rewrite the above equations in the following simple notation:
\begin{equation}
\begin{aligned}
\frac{\partial w}{\partial t} = \gamma f(w,b) +  \Delta w\\
\frac{db}{dt} = \gamma g(W,b) + d \Delta b
\end{aligned}
\end{equation}
where $d$ is eventually sent on infinity, since there is no spatial time lag for reaching the equilibrium value of $b$, and $b$ is always spatially uniform. 
After linearization around the fixed point for small $|\textbf{w}|$:
\[
\textbf{w} = 
\begin{pmatrix}
w-w^*\\
b-b^*
\end{pmatrix} \ ,
\] 
these two equations can be written in vector form as follows
\begin{equation}{\label{RD_full}}
\frac{\partial {\bf w}}{\partial t} = \gamma A \textbf{w} + D\Delta \textbf{w} \ , \quad \text{where} \quad D = \begin{pmatrix}
1 & 0 \\
0 & d
\end{pmatrix}
\quad \text{and} \quad A = \begin{pmatrix}
f_w & f_b\\
g_w & g_b
\end{pmatrix}_{(w^*,b*)} = 
\begin{pmatrix}
2 + 3(w^*)^2 - b^* & -w^*\\
8(w^*)^3L & 4(1-b^*)
\end{pmatrix}
\end{equation} 
are, respectively, the diffusion and stability matrix.

We impose first that  the uniform fixed point should be stable in the absence of any spatial variation, as we demande that the instability solely comes from spatial interactions. Keeping the non-spatial part of the equation:
\begin{equation}
\frac{\partial {\bf w}}{\partial t} = \gamma A \textbf{w}
\end{equation}
We look for solutions of the form $\textbf{w} = e^{\lambda t}\, {\bf w}_0$. Stability requires that $\text{Re}(\lambda)$ be $<0$, that is,
\begin{equation}
\begin{aligned}
\text{tr} A = f_w + g_b <0 \\
\det A = f_w g_b - f_b g_w >0
\end{aligned}
\end{equation}
It is easy to check that these general conditions, once applied to the derivatives of $f$ and $g$ listed in (\ref{RD_full}), are satisfied as soon as $L>1$.

We then ask for the existence of ann instability resulting from the spatial part of the equation. As the Laplacian operator is translation invariant we look for a solution to the reaction-diffusion system (\ref{RD_full}) that can be decomposed onto Fourier wave planes of momentum $k$ multiple of $2\pi/L$ due to periodic boundary conditions:
\begin{equation}\label{RD_time}
\textbf{w}(x,t) = \sum_{k}c_k\,  e^{\lambda t}\, \textbf{W}_{k}(x)\ ,
\end{equation}
where the constants $c_k$ are determined by a Fourier expansion of the initial conditions in terms of $\textbf{W}_k(x)$ and $\lambda$ is the eigenvalue that determines the temporal growth of the instability. Inserting (\ref{RD_time}) into (\ref{RD_full}), we get for each $k$,
\begin{equation}
\lambda \textbf{W}_k= \gamma \, A \textbf{W}_k - k^2\, D \textbf{W}_k\ .
\end{equation}
Hence, $\lambda$ is the root of the following characteristic polynomial:
\begin{equation}
\det \big( {\lambda\, I - \gamma \, A + k^2\,D} \big)= 0 \ .
\end{equation}
For the uniform steady state ($w^*,b^*$) to be unstable against spatial fluctuations, we require Re$( \lambda)>0$ for some $k\neq 0$. The conditions for this can be easily worked out, with the result
\begin{equation}{\label{cond3}}
d\, f_w+ g_b >  0 \ , 
\end{equation}
and
\begin{equation}{\label{cond4}}
(d\, f_w + g_b)^2 - 4d(f_w\, g_b - f_b\, g_w) >0\ .
\end{equation}
Condition (\ref{cond4}) is always satisfied since $d\to +\infty$. To check (\ref{cond3}) we have to evaluate the coefficient of $d$, which is $f_w$:
\begin{equation}
f_w= 2 + 3(w^*)^2 - b^* =\frac{4}{L-1} > 0  \ .
\end{equation}
Hence, this condition is satisfied as soon as $L>1$.

\bibliography{bibliography.bib}

\begin{thebibliography}{99}

\bibitem{manifold}
Y. Bengio, A. Courvile, P. Vincent. Representation Learning: A Review and New Perspectives, {\em IEEE Transactions on Pattern Analysis and Machine Intelligence} {\bf 35} , 1798-1828 (2013)

\bibitem{Laughlin}
S. Laughlin. A simple coding procedure enhances a neuron's information capacity. {\em Z. Naturforsch} {\bf 36}, 910-912 (1981)

\bibitem{atick1992could}
J.J. Atick. Could information theory provide an ecological theory of sensory processing? {\em Network: Computation in neural systems} {\bf 3}, 213-251 (1992)

\bibitem{Vijay}
P. Garrigan, C.P. Ratliff, J.M. Klein, P. Sterling, D.H. Brainard, V. Balasubramanian. Design of a trichromatic cone array. {\em PLoS Computational Biology} {\bf 6}, e1000677
(2010)

\bibitem{Tibi}
T. Tesileanu, S. Cocco, R. Monasson, V. Balasubramanian. Adaptation of olfactory receptor abundances for efficient coding . {\em eLife} {\bf 8}, e39279 (2019)

\bibitem{gerstner}
C.S.N. Brito, W. Gerstner. Nonlinear Hebbian learning as a unifying principle in receptive field formation. {\em PLoS Computational Biology} {\bf 12}, e1005070 (2016)

\bibitem{sparse}
B.A. Olshausen, D.J. Field.  Emergence of simple-cell receptive field properties by learning a sparse code for natural images. {\em Nature} {\bf 381}, 607?609 (1996)

\bibitem{hyvarinen2000independent}
A. Hyv{\"a}rinen, E. Oja. Independent component analysis: algorithms and applications. {\em Neural networks} {\bf 13}, 411--430 (2000)

\bibitem{monasson93}
R. Monasson. Storage of spatially correlated patterns in auto-associative memories. {\em J. Physique I} {\bf 3}, 1141 (1993)

\bibitem{sengupta}
A.M. Sengupta, M. Tepper, C. Pehlevan, A. Genkin, D.B. Chklovskii. Manifold-tiling Localized Receptive Fields are Optimal in Similarity-preserving Neural Networks. {\em NeurIPS} 7080-7090 (2018)

\bibitem{benna}
M.K. Benna, S. Fusi. Are place cells just memory cells? Memory compression leads to spatial tuning and history dependence. bioRxiv 624239 (2019).

\bibitem{battista19}
A. Battista, R. Monasson. Capacity-resolution trade-off in the optimal learning of multiple low-dimensional manifolds by attractor neural networks. arXiv:1910.05941
(2019) 

\bibitem{mehtaschwab}
P. Mehta, D.J. Schwab. An exact mapping between the Variational Renormalization Group and Deep Learning. arXiv:1410.3831 (2014) 

\bibitem{ringel}
M. Koch-Janusz, Z. Ringel. Mutual information, neural networks and the renormalization group. {\em Nature Physics} {\bf 14}, 578-582 (2018)

\bibitem{watkin}
T. L. H. Watkin, J-P Nadal. Optimal unsupervised learning. {\em J. Phys. A} {\bf 27}, 1899 (1994)

\bibitem{reimann}
P. Reimann, C. Van den Broeck. Learning by examples from a nonuniform distribution. {\em Phys. Rev. E} {\bf 53}, 3989--3998 (1996)

\bibitem{bbp}
J. Baik, G. Ben Arous, S. Pech\'e. Phase transition of the largest eigenvalue for nonnull complex sample covariance matrices. {\em Annals of Probability} {\bf 33}, 1643--1697 (2005)

\bibitem{barra2017phase}
A. Barra, G. Genovese, P.  Sollich, D. Tantari. Phase transitions in Restricted Boltzmann Machines with generic priors. {\em Phys. Rev. E} {\bf 96}, 042156 (2017)

\bibitem{le2008representational}
N. Le Roux, Y. Bengio. Representational power of restricted Boltzmann machines and deep belief networks. {\em Neural computation} {\bf 20}, 1631--1649 (2008)

\bibitem{tubiana2017emergence}
J. Tubiana, R. Monasson. Emergence of compositional representations in restricted Boltzmann machines. {\em Phys. Rev. Lett.} {\bf 118}, 138301 (2017)

\bibitem{tieleman2008training}
T. Tieleman. Training restricted Boltzmann machines using approximations to the likelihood gradient. {\em Proceedings of the 25th international conference on Machine learning}, 1064--1071 (2008)

\bibitem{fischer2015}
A. Fischer, C. Igel. An introduction to restricted Boltzmann machines. In {\em Iberoamerican Congress on Pattern Recognition}, 14--36, Springer, Buenos Aires, Argentina (2012)

\bibitem{jastrzkebski2017three}
S. Jastrzkebski, Z. Kenton, D. Arpit, N.  Ballas, A. Fischer, Y. Bengio, A. Storkey. Three factors influencing minima in SGD. arXiv:1711.04623 (2017).

\bibitem{hochreiter1997flat}
S.   Hochreiter, J.  Schmidhuber. Flat minima. {\em Neural Computation} {\bf 9}, 1--42 (1997). 

\bibitem{keskar2016large}
N.S. Keskar, D. Mudigere, J. Nocedal, M. Smelyanskiy, P.T.P. Tang. On large-batch training for deep learning: Generalization gap and sharp minima. 
 arXiv:1609.04836 (2016)

\bibitem{chaudhari2016entropy}
P.  Chaudhari, A. Choromanska, S. Soatto, Y. LeCun, C. Baldassi, C. Borgs, J. Chayes, L. Sagun, R. Zecchina. Entropy-SGD: Biasing gradient descent into wide valleys.
arXiv:1611.01838 (2016)

\bibitem{bottou2008tradeoffs}
L. Bottou, O. Bousquet. The tradeoffs of large scale learning. {\em Advances in neural information processing systems}, 161--168 (2008)

\bibitem{smith2017bayesian}
S.L. Smith, Q.V. Le.  A bayesian perspective on generalization and stochastic gradient descent.  arXiv:1710.06451 (2017)

\bibitem{decelle}
A. Decelle, G. Fissore, C. Furtlehner. Spectral dynamics of learning in restricted Boltzmann machines. {\em Europhys. Lett.} {\bf 119}, 60001 (2017)

\bibitem{book1}
J.D. Murray. Mathematical Biology. II. Spatial Models and Biomedical Applications. In {\em Interdisciplinary Applied Mathematics, vol. 18}. Springer, New York (2003)

\bibitem{amari}
S. Amari. Dynamics of pattern formation in lateral-inhibition type neural fields. {\em Bio. Cyber.} {\bf 27}, 77-87 (1977)

\bibitem{tsodyks}
M. Tsodyks, T. Sejnowski. Associative Memory and Hippocampal Place Cells.
{\em Int. J. Neur. Syst.} {\bf  6}, 81-86 (1995)

\bibitem{sompo}
B. Ben-Yishai, R. Bar-Or, H. Sompolinsky. Theory of orientation tuning in visual cortex. {\em Proc. Natl. Acad. Sci. } {\bf 92}, 3844-48 (1995)

\bibitem{treves}
A. Treves. {\em private communication.} (2019)

\bibitem{Casali19}
G. Casali, D. Bush, K. Jeffery. Altered neural odometry in the vertical dimension. {\em Proc. Natl. Acad. Sci.} {\bf 116}, 4632-4636 (2019)

\bibitem{Hayman11}
R. Hayman, M. Verriotis, A. Jovalekic, A.A. Fenton, K. Jeffery. Anisotropic encoding of three-dimensional space by place cells and grid cells. {\em Nat. Neurosci.} {\bf 14}, 1182-1188 (2011)

\bibitem{vrokeefe}
G. Chen, J.A. King, N. Burgess, J. O'Keefe. How vision and movement combine in the hippocampal place code. {\em Proc. Natl. Acad. Sci.} {\bf 110}, 378--383 (2010)

\end{thebibliography}

\end{document}